\documentclass[fleqn,usenatbib]{mnras}

\usepackage{newtxtext,newtxmath}

\usepackage[T1]{fontenc}

\DeclareRobustCommand{\VAN}[3]{#2}
\let\VANthebibliography\thebibliography
\def\thebibliography{\DeclareRobustCommand{\VAN}[3]{##3}\VANthebibliography}

\usepackage{graphicx}	
\usepackage{amsmath}	
\usepackage{dblfloatfix}
\usepackage{caption}
\usepackage{nccmath}

\usepackage[justification=centering]{caption}

\title[]{Most of the cool CGM of star-forming galaxies is not produced by supernova feedback}

\author[A. Afruni et al.]{
Andrea Afruni,$^{1}$\thanks{E-mail: afruni@astro.rug.nl}
Filippo Fraternali,$^{1}$
Gabriele Pezzulli$^{1,2}$
\\
$^{1}$Kapteyn Astronomical Institute, University of Groningen,Landleven 12, 9747 AD Groningen, The Netherlands\\
$^{2}$ Department of Physics, ETH Z\"{u}rich, Wolfgang-Pauli-Strasse 27, 8093 Z\"{u}rich, Switzerland\\
}

\date{Accepted XXX. Received YYY; in original form ZZZ}

\pubyear{2020}

\begin{document}
\label{firstpage}
\pagerange{\pageref{firstpage}--\pageref{lastpage}}
\maketitle

\begin{abstract}
The characterization of the large amount of gas residing in the galaxy halos, the so called circumgalactic medium (CGM), is crucial to understand galaxy evolution across cosmic time. We focus here on the the cool ($T\sim10^4$ K) phase of this medium around star-forming galaxies in the local universe, whose properties and dynamics are poorly understood.
We developed semi-analytical parametric models to describe the cool CGM as an outflow of gas clouds from the central galaxy, as a result of supernova explosions in the disc (galactic wind). The cloud motion is driven by the galaxy gravitational pull and by the interactions with the hot ($T\sim10^6$ K) coronal gas.
Through a bayesian analysis, we compare the predictions of our models with the data of the COS-Halos and COS-GASS surveys, which provide accurate kinematic information of the cool CGM around more than 40 low-redshift star-forming galaxies, probing distances up to the galaxy virial radii.
Our findings clearly show that a supernova-driven outflow model is not suitable to describe the dynamics of the cool circumgalactic gas. Indeed, to reproduce the data, we need extreme scenarios, with initial outflow velocities and mass loading factors that would lead to unphysically high energy coupling from the supernovae to the gas and with supernova efficiencies largely exceeding unity. This strongly suggests that, since the outflows cannot reproduce most of the cool gas absorbers, the latter are likely the result of cosmological inflow in the outer galaxy halos, in analogy to what we have previously found for early-type galaxies.
\end{abstract}




\section{Introduction}\label{intro}
The perfect laboratory to study and understand how galaxies evolve through cosmic time is the ionized gas that resides in the region between them and the intergalactic medium (IGM), the so called circumgalactic medium (CGM). This medium is observed, at very different temperatures, both around our Milky Way and almost ubiquitously around external galaxies \citep[e.g.][]{anderson11,werk13,miller15,tumlinson17}. From a theoretical point of view, the halos of galaxies are expected, depending on their mass \cite[e.g.][]{birnboim03}, to be filled with hot gas at about the virial temperature (generally called corona, predicted decades ago by cosmological models, see \citealt{white78}), and with colder gas likely distributed along filaments, that can either penetrate to the halo central regions \citep{dekel09} or evaporate into the hot corona \citep{nelson13}.\\ 
Although the halos of galaxies with different masses and morphologies show the presence of very large amounts of  cool ( $T\sim 10^4\ \rm{K}$) CGM absorbers \citep[e.g.][]{thom12,stocke13,bordoloi14,heckman17,zahedy19} there is still much debate on the general dynamics of these clouds and on their possible origins, which might also depend on the galaxy type. For passive early-type objects, given the absence of activity in the center, the cool clouds probably originate either from the inflow of external intergalactic medium \citep{afruni19} or from the condensation of the hot coronal gas \citep{voit18,nelson20}. Also for star-forming galaxies, observations of this cool gas had in some cases been interpreted as clouds falling towards the galaxy, presumably feeding its star formation \citep[e.g.][]{bouche13,borthakur15}, as expected from theoretical models. For these star-forming objects however, the central galaxy is believed to have an active role in the formation and regulation of the cool CGM. Over the years multiphase outflows have been observed in the central regions of both dwarfs \citep[e.g.][]{mcquinn19} and $L_{\ast}$ spiral galaxies \citep[e.g.][]{veilleux05,martin12,rubin14,concas19}, with claims of these winds being part of large-scale galactic outflows, extending till several tens of kpc from the center \citep[e.g.][]{schroetter19}. It is not clear however whether these ionized outflows are powered by star formation and, if so, what is their impact on the surrounding CGM (see \citealt{martin09,borthakur13}, who studied the properties of the circumgalactic gas of starburst galaxies).\\
Generally, at the typical scales of the CGM ($\sim100$ kpc), it is hard to distinguish whether the cool gas is outflowing from or inflowing to the central object, given the limited information coming from the observations. Despite having now quite some evidence of CGM emission at high redshift \citep[e.g.][and references therein]{cantalupo14,farina19}, observations of the cool gas around galaxies in the local universe are primarily in absorption and consist of one single line of sight for each galaxy \citep[see][and references therein]{tumlinson17}, with very few examples of observations in emission \citep[e.g.][]{burchett20}, sometimes using stacking techniques \citep[see][]{zhang18}. There are therefore very few constraints on the physical position of the cool clouds and on their intrinsic dynamics.\\ Different studies, focused on metal UV absorption lines \citep{kacprzak12,martin12,schroetter19,veilleux20} have found a segregation of absorbers along the galaxy minor and major axes, which would hint towards a bi-conical outflow scenario, with accretion along the disc plane. The same feature is however not observed in other samples where the absorptions are more uniformly distributed \citep[see][]{borthakur15,pointon19} and therefore more statistics is needed in order to draw any conclusion on the gas origin or dynamics.\\
One key ingredient to disentangle between inflow or outflow motion is the gas metallicity: clouds inflowing from the IGM are expected to have very low metallicities \citep{lehner16}, while gas originated from the supernova triggered winds will transport a larger amount of metals. Deriving this gas property is however not trivial, since it requires photo-ionization modeling with multiple underlying assumptions, especially on the gas ionization factor \citep{werk14,wotta16}. Generally, both low and high metallicity absorbers are observed \citep[e.g.][]{prochaska17}, without however a clear dependence between the metallicity and the azimuthal position of the absorbers \citep{peroux16,kacprzak19,pointon19}, as we would expect instead from the scenario of bi-conical outflows plus longitudinal accretion.\\
The study of the CGM from a theoretical point of view is mostly based on hydrodynamical simulations that can trace the whole amount of gas inside a single galaxy halo, either with 'zoom-in' simulations of cosmological suites \citep[e.g.][]{muratov17,pillepich18,oppenheimer18,rahmati18} or with idealized ad-hoc simulations of a single galaxy \citep[e.g.][]{fielding17}. In both approaches, central winds seem to play an important role in defining the CGM properties and, while part of the cool gas is coming from the accretion of pristine gas, a significant fraction is either outflowing from the center or recycling back in the form of metal enriched clouds, after being previously ejected \citep[e.g.][]{ford14,ford16,alcazar17,oppenheimer18}.\\
However, these simulations often have to rely on subgrid models to treat the physics of stellar feedback, that make the predictions of the circumgalactic gas properties not completely reliable and in many cases, like the predicted gas metallicity distributions, not in agreement with the observations \citep[see][]{wotta19}. Moreover, the main limitation is given by the resolution, that can reach at best a kpc-scale \citep[e.g.][]{vandevoort19}, which is not high enough to properly trace and resolve the clumpy cool circumgalactic gas in the galaxy halos. The general properties and structure of this gas, as well as its kinematics \citep{peeples19}, are indeed dependent on the resolution of the simulations \citep{vandevoort19}, without clear signs of convergence. Many high-resolution hydrodynamical simulations \citep{armillotta16,schneider18,mccourt18,gronnow18,gronke18,fielding20} have been focusing on the interactions between hot and cold gas, finding that at least a pc-scale resolution is necessary to resolve the instabilities developing at the cloud/corona interface and therefore to properly describe the evolution of the cold clouds, resolution that is far from being achievable in simulations of the entire galaxy halo.\\
To overcome the issues related to simulations we developed in this work semi-analytic parametric models. The analytic approach is rarely used to understand the circumgalactic medium \citep[e.g.][]{stern16,lan18,afruni19} and the few works done so far have very different characteristics and goals between each other. However, the ability of an analytical study to describe the whole circumgalactic medium distribution within the galaxy halo, with straight-forward assumptions on the gas physics and origin, is key to understanding the observational data and draw conclusions on the CGM properties and dynamics. In our previous work \citep{afruni19}, with a comparison of our model predictions with kinematic data from \cite{zahedy19}, we have shown that the cool circumgalactic gas of early-type galaxies is consistent with an inflow of clouds coming from the cosmological accretion of gas onto the galaxy halos.\\
Here, we will compare the predictions of our models with the observations of the COS-Halos and COS-GASS surveys \citep[see Section~\ref{Observations},][]{werk13,borthakur15}, that provide very accurate kinematic data to constrain our models, and we will focus in particular on a sample of star-forming galaxies. This direct comparison with high resolution data will help us interpret the CGM properties for galaxies similar to our Milky Way. In particular, we describe the cool circumgalactic gas as an outflow of clouds generated from the supernova explosions in the central galaxies, taking into account the combined effect on the cloud orbits of the gravitational pull of the galaxies and of the interactions between the clouds and the hot coronal gas. With our analysis, we aim to gain more insight on the role of galactic supernova-driven outflows in the dynamics and origin of this cool gas.\\
This paper is organized as follows: in Section~\ref{Observations} we show the sample of galaxies and the absorption kinematic data that we will use in this work; in Section~\ref{model} we describe how we built our semi-analytic models; in Sections~\ref{results} and \ref{discussion} we report our results and we discuss the implications of our findings, while in Section~\ref{conclusions} we summarize our work and conclusions.
\section{Galaxy sample and data}\label{Observations}
\begin{figure}
    \includegraphics[width=1\linewidth]{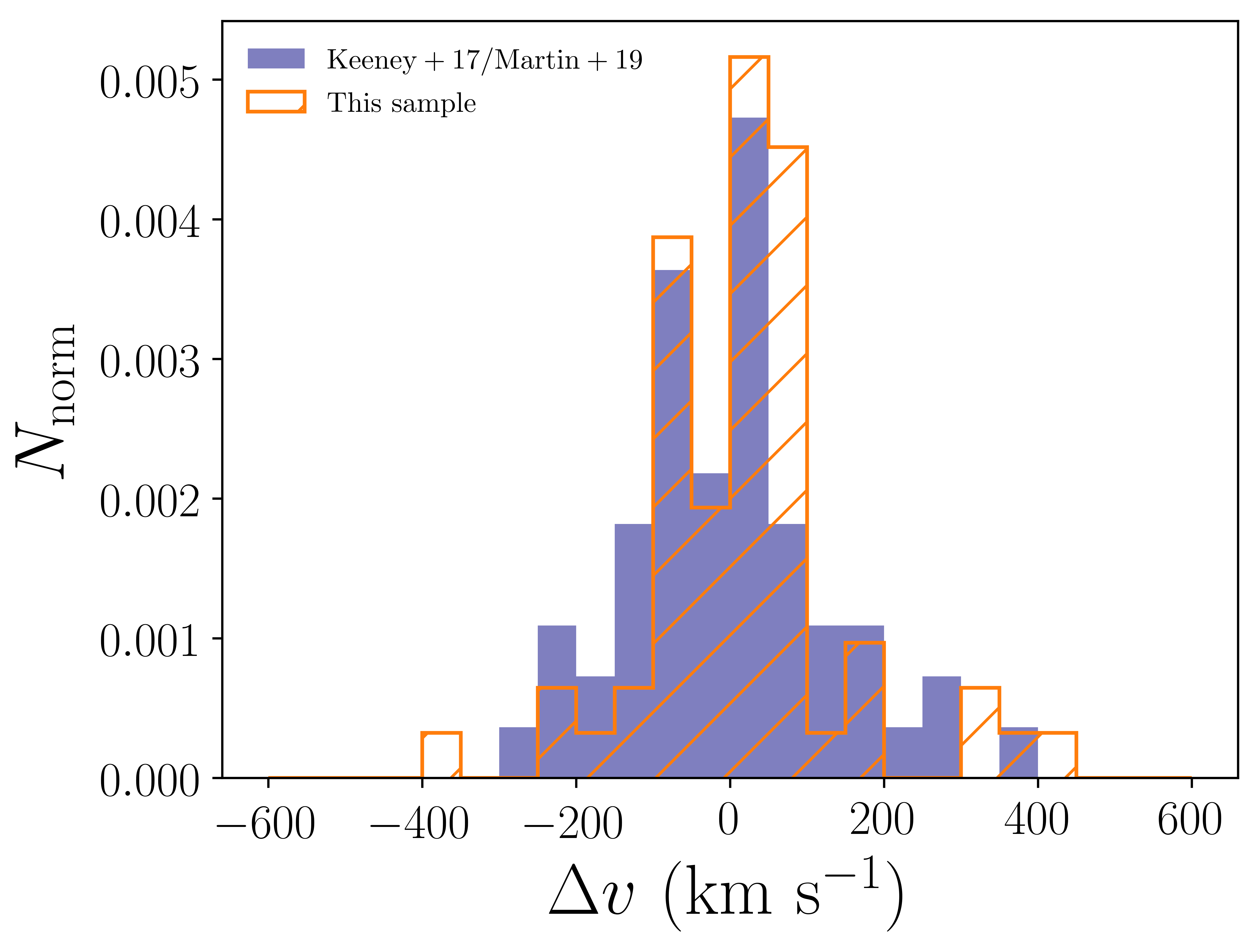}
    \caption{Orange-hatched histogram: velocity distribution of all the 62 Ly$\alpha$ components identified in the 41 QSO spectra in our sample. Purple histogram: velocity distribution of the cool CGM for a subsample of star-forming galaxies with stellar masses consistent with our main sample,
    drawn from \citep{keeney17,martin19}.}
      \label{fig:velobs}%
    \end{figure}

\begin{table*}
\begin{center}
\begin{tabular}{*{9}{c}}
(1)&(2)&(3)&(4)&(5)&(6)&(7)&(8)&(9)\\
\hline  
\hline
Galaxy ID  &      z  & $\log(M_{\ast}/M_{\odot})$ & $\rm{SFR}\ (M_{\odot}\ \rm{yr}^{-1})$ & $n_{\rm{comp}}$ &  $x_{\rm{los}}$ (kpc)& $y_{\rm{los}}$ (kpc)& $R_{\rm{d}}$ (kpc)& $i$ ($^{\circ}$)  \\
\hline
3936 &        0.0441 &  10.1 & 3.98 &  2 &  38 &  99 & 2.6 & 65   \\
20042 &        0.0468 & 10.0 & 2.51 & 1 & -130 &    -146 &  1.6 & 49  \\
8634 &       0.0464 & 10.1 & 0.20 & 1 & -97 &   -34 &    0.6 & 25  \\
23457 &       0.0354 & 10.1 & 0.25 & 1 & 76 &    -158 &    2.8 &  82 \\
29871 & 0.0342 & 10.2 & 3.16 & 1 & -230 & -27 &          3.0 & 61  \\
38018 &       0.0297 & 10.1 &0.40& 1 & -32 &   -156&    2.9 & 84  \\
42191  &    0.0320 & 10.1 &2.00&  1 & -232 &   72 &   0.7 & 49  \\
41869 & 0.0414 & 10.1 &3.16& 3 & 83 &  103 &    2.0 & 68  \\
$170\_9$ &  0.3557 & 10.0 &3.04& 2 & -33 & -33 &    1.9 & 20  \\
$274\_6$ &  0.0252 & 9.9 &0.64& 2 & 13 &   31 &    3.2 & 36  \\
$359\_16$  & 0.1661 & 10.2 &1.37&  1 & -13 &   44 &  3.0 & 38  \\
$236\_14$ & 0.2467 & 10.0 &5.68& 2 & -25 &  58 &   3.7 &  32  \\
$168\_7$ &  0.3185 & 10.2 &3.42& 3 & 29 & -9 &    4.5 & 36  \\
$289\_28$ & 0.1924 & 10.1 &1.99& 2 & -43 & 92 &    2.1 & 37  \\
$126\_21$ & 0.2623 & 10.1 &5.56& 2 & -74 & -65 &     3.4 & 44  \\
$232\_33$ & 0.2176 & 10.1 &2.60& 2 & -88 & 96 &    2.3 & 48   \\
$88\_11$ &  0.1893 & 10.1 &4.18& 1 & 27 & -31 &   3.7 & 44   \\
\hline
8096 &      0.0345 & 10.3 &1.58& 1 & -83 & 158 &    2.1 & 59   \\
32907 & 0.0349 & 10.5 &0.63& - & 204 & -78 &    2.8 & 80   \\
23419 & 0.0400 & 10.4 &2.51& 1 & -48 &  132 &    2.4 & 70  \\
49433 & 0.0458 & 10.5 &1.58& 2 & 231 & 25 &     1.9 & 40   \\
50550 & 0.0350 & 10.3 &1.99&  1 & 158 & 121 &   1.8 & 52   \\
13159 & 0.0437 & 10.4 &0.40&  2 & 100 & -27 &   1.9 & 75   \\
51025  & 0.0450 & 10.3 &0.79& 1 & -47  & 214&  2.3 & 74  \\
41743  &0.0462   & 10.5  &1.99& 2 & -57 & -218  & 2.7  & 69   \\
28365 &  0.0321   & 10.4  &6.3& 1 & 124 & -27  & 4.3  & 29   \\
$34\_36$  &  0.1427   & 10.4  &14.12& 2 & 12 & -114  & 3.7   & 49  \\
$106\_34$  & 0.2284   & 10.5  &4.52& 1 & 61 & -108  & 3.7  & 19   \\
$94\_38$ &  0.2221   & 10.5  &4.38& 4 & -204 & -13  & 4.1  & 59   \\
$349\_11$ &   0.2142   & 10.5  &0.62& 1 & 32 & 23  & 3.7  & 37   \\
$132\_30$  &0.1792   & 10.3  &11.36& 1 & 110 & -7  & 3.1  & 12  \\
\hline
55745  &0.0278   & 10.9  &3.98& 1 & -16 & -62  & 6.6  & 35   \\
22822 &  0.0270   & 10.6  &1.58& 1 & 228 & -96  & 1.8  & 64  \\
55541 &  0.0429   & 10.6  &3.16& 1 & -120 & 194  & 3.9  & 81   \\
5701 &  0.0422   & 10.7  &0.63& 1 & 141 & -141  & 2.1  & 31   \\
48604 &  0.0334   & 10.6  &0.40& 2 & -117 & -90  & 2.2  & 50  \\
48994 &  0.0322   & 10.7  &1.99& 1 & 79 & -74  & 6.7  & 86   \\
13074 &  0.0486   & 10.9  &3.16& 2 & 176 & -98  & 3.5  & 71    \\
$157\_10$  & 0.2270   & 10.7  &6.04& 3 & 33 & -12  & 3.1  & 25   \\
$97\_33$  &  0.3218   & 10.6  &7.42& 1 & 23 & -197  & 5.7  & 61  \\
$68\_12$  &  0.2024   & 10.8  &18.96& 2 & -44 & -15  & 6.8  & 30   \\
\hline
\end{tabular}
\captionsetup{justification=centering}
\caption[]{Properties of the galaxies in our sample. (1) galaxy ID; (2) redshift; (3) stellar mass; (4) star formation rate; (5) number of kinematic components identified in the quasar spectrum \cite[from][]{tumlinson13,borthakur15}; (6) and (7) x and y coordinates of the line of sight with respect to the galactic disc; (8) stellar disc scalelength; (9) inclination. }\label{tab:galprop}
\end{center}
\end{table*}
The findings of this work are obtained through the comparison of our model predictions with the observational data of the COS-Halos and COS-GASS surveys \citep{werk13,borthakur15}, which
are focused on the cool CGM around low-redshift early and late-type galaxies, over a large range of stellar masses. These observations are taken pointing the Cosmic Orign Spectrograph \citep[COS,][]{froning09} aboard the Hubble Space Telescope towards background quasars (QSOs) in the projected vicinity of the galaxies. The gas is then characterized through the analysis of the hydrogen and metal absorption lines \citep[][]{tumlinson13,werk14,borthakur15,borthakur16} in the QSO spectra. One single galaxy is associated to each QSO and the impact parameters (projected distance between the central galaxy and the line of sight of the quasar) lie in the range between 10 kpc and 250 kpc from the central object, probing the circumgalactic gas from the center up to the galaxy virial radius. For a detailed description of the two surveys, see the COS-Halos and COS-GASS papers.\\
The purpose of this work is to derive the properties of the CGM of typical star-forming galaxies and the impact of supernova-driven galactic outflows on the cool gas dynamics and origin. 
To this end, we selected only a subsample of 41 disc galaxies, that satisfy the two criteria of being star-forming ($\rm{sSFR}>10^{-11}\ \rm{yr}^{-1}$) and having a stellar mass $10^{10}\lesssim M_{\ast}/M_{\odot}<10^{11}$. With this selection we therefore excluded dwarf galaxies and massive passive galaxies \citep[but see][]{afruni19}, where the cool CGM could have different origins or dynamics (see Section~\ref{discussion}).\\
As a comparison to the predictions of our models, we will use in this work the kinematic information provided by the two surveys. For both studies, UV absorption lines of both low-ionization metals and neutral hydrogen are identified in the QSO spectra through a Voigt profile fitting analysis \citep[][]{werk13,tumlinson13,borthakur15,borthakur16} in a spectral window that goes from $-600$ to $+600\ \rm{km}\ \rm{s}^{-1}$ from the systemic velocity of the central galaxy, with a kinematic resolution of about $18\ \rm{km}\ \rm{s}^{-1}$. Cool gas is observed in all but one spectra in our sample\footnote{For the non-detection, \cite{borthakur15} report an equivalent width equal to three times the noise in the spectrum in the vicinity of the expected transition.}.\\ 
For consistency, we decided to focus only on one tracer and therefore to use in this work only the data concerning the hydrogen $\rm{Ly}\alpha$ lines. This line (similarly to the metal ones) is observed in the same spectrum with different velocities, identifying different kinematic components. The presence in the spectra of multiple-component absorptions implies that the cool CGM is not a homogeneous and uniform layer of gas, but rather a composition of different clouds moving throughout the halos with a complex kinematics, a common feature found by many different studies \citep[e.g. ][]{bordoloi14,werk16,stern16,keeney17,zahedy19}. This will be a fundamental assumption for our models. The total number of $\rm{Ly}\alpha$ components found around the 41 galaxies of our sample is 62 and in Table~\ref{tab:galprop} we report the number of components for each galaxy-QSO pair. The average number of components per line of sight is $1.5$.
   \begin{figure}
   \includegraphics[width=1\linewidth]{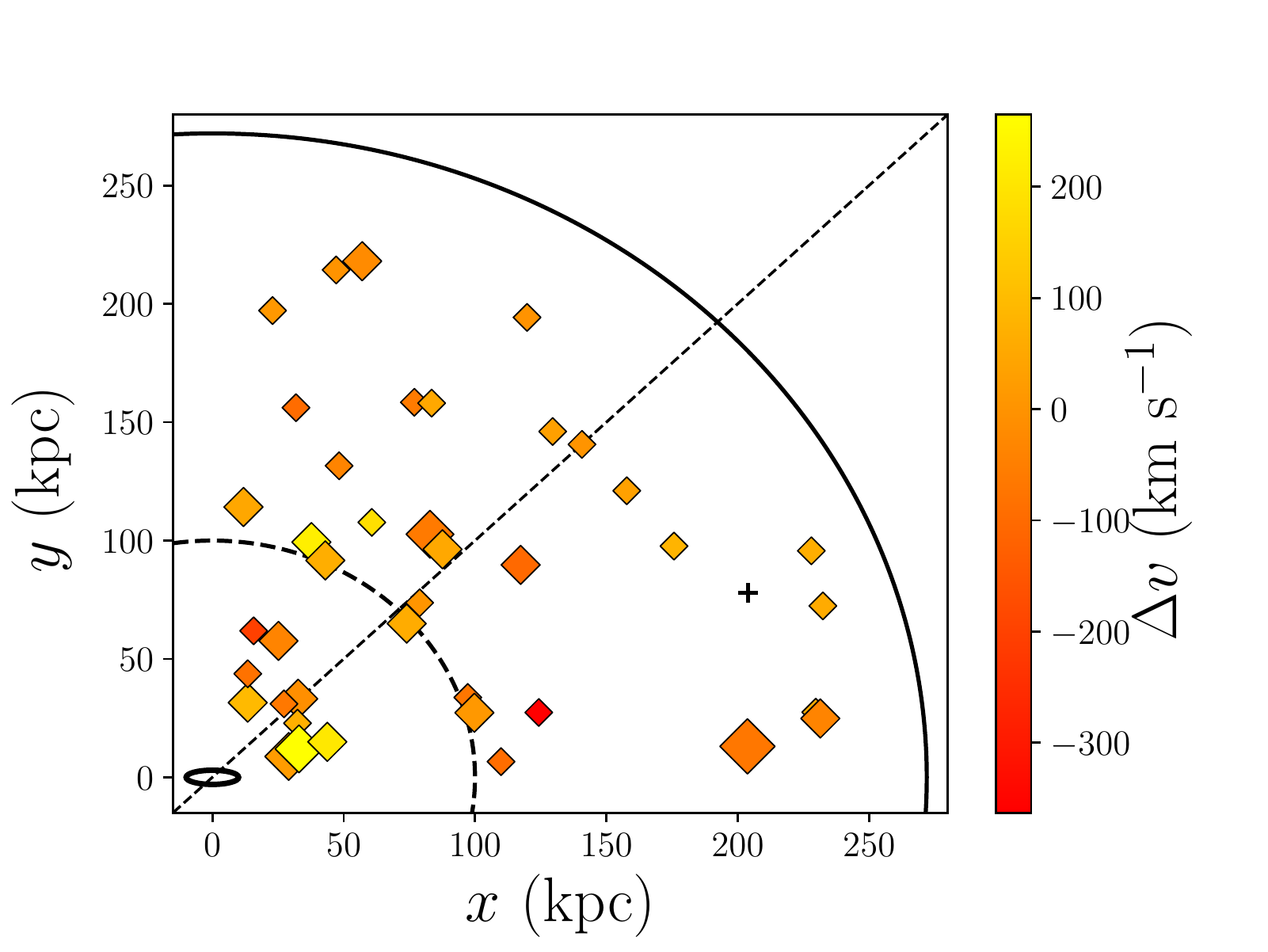}
   \caption{Plane of the observations for our subsample of galaxies taken from the COS-Halos and COS-GASS surveys. The ellipse at the bottom left corner represents the central disc galaxy, while the symbols depict the QSO lines of sight, placed at their corresponding distance from the central object, with the black cross representing the non-detection. The colorbar shows the average velocity of the cool CGM found for each sightline, while the size of the symbols is related to the number of components identified in each spectrum. The black solid curve represents the median virial radius of our galaxy sample (272 kpc), while the dashed curve represents a radius of 100 kpc. The dashed straight line depicts instead the bisector of the plane.}
              \label{fig:Obslosplane}%
    \end{figure}\\
All the properties of our sample relevant to our analysis are reported in Table~\ref{tab:galprop} and are retrieved directly from \cite{tumlinson13} and \cite{borthakur15}, except for the geometric parameters (coordinates of the sightlines with respect to the galactic discs, disclengths and disc inclinations), which are obtained performing a fit for each galaxy using the software GALFIT \citep{peng10}. The details of the fitting procedure are explained in Appendix~\ref{galfit}.
In Figure~\ref{fig:Obslosplane} we report the absolute values of the positions of all the lines of sight, together as in one single halo, with the galaxy at the center and the x-axis and y-axis corresponding respectively to the major and minor axis of the projected galaxy disc. The exact position of each line of sight with respect to the central object is reported in Table~\ref{tab:galprop} and was inferred through the GALFIT analysis (see Appendix~\ref{galfit}). 
The black solid curve depicts the median virial radius (see Section~\ref{model}) of the 41 galaxies in our sample, equal to $r_{\rm{vir}}=272$ kpc. The size of the symbols in Figure~\ref{fig:Obslosplane} represents the number of components found for each sightline, which varies from one to four (the cross represents instead the only non-detection), while the different colours represent the average Ly$\alpha$ absorption velocities with respect to the systemic velocity of the central galaxies.\\
From Figure~\ref{fig:Obslosplane} it is therefore clear how the observations give us information spanning the entire extension of the galaxy halos, with the limitation however that each object has only one sightline associated to it. It is also important to note that, contrary to the claims of other surveys \citep[e.g.][]{schroetter19,martin19}, where the cool gas absorbers seem to be found primarily along the galaxy major/minor axis, this dataset does not show any evidence of a preferential orientation for the absorption of the cool CGM, whose detections are uniformly distributed throughout the halo. We will see in Section~\ref{results} how this feature influences the results of our analysis.\\ 
In Figure~\ref{fig:velobs} we report instead in orange the velocity distributions of all the detected components, with values ranging approximately from -400 to 400 km $\rm{s}^{-1}$. As a comparison, we also show in purple the velocity distribution of the cool CGM around galaxies selected from the surveys of \cite{keeney17} and \cite{martin19} using the same two criteria on the stellar mass and star formation rate previously used for our sample. These surveys have features similar to the ones of COS-Halos and COS-GASS, but will not be directly used in this work as a constraint for our models\footnote{\cite{martin19} observed the cool gas through MgII, while we focus here only on the hydrogen lines. The galaxies from \cite{keeney17} have instead less strict conditions for isolation with respect to the COS-Halos and COS-GASS galaxies, therefore their cool CGM is more likely to be contaminated by other objects.}. We can see however from Figure~\ref{fig:velobs} how the kinematics of the absorbers of our sample is representative of the one found by different studies, which justifies the choice of these two surveys as our fiducial dataset.
    \\
Figures~\ref{fig:Obslosplane} and \ref{fig:velobs} give us an overview of all the information derived from the COS observations. The cool CGM kinematics is derived throughout the whole galaxy halos, combining both the velocity of the absorbers, their number and their projected position with respect to the central galaxy. The aim of this work is to reproduce, through dynamically motivated models, all the observed velocity components at their distance from the galaxy.
\begin{table*}
\begin{center}
\begin{tabular}{ccc|ccccccc}
(1)&(2)&(3)&(4)&(5)&(6)&(7)&(8)&(9)&(10)\\
\hline  
\hline
\textbf{Model name}  &  $M_{\ast}$ range  & $n_{\rm{obj}}$ & $\log(M_{\ast}/M_{\odot})$ & z & $\rm{SFR}$ & $R_{\rm{d}}$ & $\log(M_{\rm{vir}}/M_{\odot})$ & $r_{\rm{vir}}$ & $T_{\rm{vir}}$  \\
               &     & & && $(M_{\odot}\ \rm{yr}^{-1})$ & (kpc) & &(kpc)& $(10^6\ \rm{K})$ \\
\hline
\textbf{Gal1} & $9.9\leq \log(M_{\ast}/M_{\odot})<10.3$ & 17 &10.1&0.1661&2.60 &2.53 &11.9&228&0.58\\
\textbf{Gal2} & $10.3\leq \log(M_{\ast}/M_{\odot})<10.6$ & 14 &10.4&0.0454& 1.99&2.65&12.1&286&0.74\\
\textbf{Gal3} &$10.6 \leq \log(M_{\ast}/M_{\odot})< 11.0$& 10 &10.7&0.0425 & 3.16 & 3.82 &12.3&331&0.98\\
\hline
\end{tabular}
\end{center}
\captionsetup{justification=centering}
\caption[]{Properties of the 3 galaxy models described in Section~\ref{init}. (1) Model name; (2) range in stellar mass; (3) number of galaxies per subsample; (4) median stellar mass; (5) median redshift; (6) median star formation rate; (7) median stellar disc length; (8), (9) and (10) median galaxy virial mass, radius and temperature \citep[see text and][]{afruni19}.}\label{tab:modprop}
\end{table*}
\section{Model}\label{model}
We mentioned in Section~\ref{Observations} that the basic assumption that underlines our modeling is that the cool circumgalactic gas is composed of different clouds. We modelled the dynamics of these clouds taking into account the gravitational potential of the galactic disc and the virial halo and the interactions of these cool absorbers with the hot pre-existing CGM. To this end, we used the publicly available python package GALPY \citep{bovy15}, which allows to perform a 2-dimensional orbit integration within an arbitrary potential (we refer for more details on GALPY to the work of \cite{bovy15}). To develop our models, we implemented a modification in the code that takes into account the drag force acted by the hot corona, that strongly modifies the cloud motion (see Section~\ref{hydro}).
In this Section we describe how we built our parametric dynamical models for the cool CGM clouds ejected by star-forming galaxies and how we compare, through a bayesian analysis of the parameter space, the predictions of our models with the COS observations of our galaxy sample. 
\subsection{Outflow of cool CGM clouds}\label{outflow}
In this paper we investigate the scenario where the cool clouds are part of gas outflows (galactic wind) coming from the central galaxies, originated by the feedback from supernova explosions in the disc. As already introduced in Section~\ref{intro}, we model the outflow motion of the cool gas only, neglecting the effects of the hot wind, that we will discuss in Section~\ref{discussion}.
\subsubsection{Galaxy potential}\label{gravity}
In order to describe the motion of the cool clouds, we need first to assume a gravitational potential, that will pull the outflowing clouds back towards the central galaxy and, in the absence of hydrodynamical effects, determine the cloud orbits. 
We used an axisymmetric choice of the total potential, composed of two different components, the potentials of a razor-thin disc for the galaxy and of a dark matter halo described by a Navarro Frenk White profile \citep[NFW,][]{nfw96}, whose density distributions are, respectively
\begin{ceqn}
\begin{equation}\label{eq:potDisc}
\rho(R,z)=\Sigma_{\rm{d},0} \exp{(-R/R_{\rm{d}})}\ \delta(z)
\end{equation}
\end{ceqn}
\begin{ceqn}
\begin{equation}\label{eq:potNFW}
\rho(r)=\frac{\rho_{0}}{\frac{r}{r_{\rm{s}}} \left(  1+\frac{r}{r_{\rm{s}}} \right)^2}\ ,
\end{equation}
\end{ceqn}\\
where $\Sigma_{\rm{d},0}=M_{\ast}/(2\pi R^2)$ and $R_{\rm{d}}$ are the central surface density of the disc and its scale length (the latter obtained from the GALFIT analysis, see Appendix~\ref{galfit}), while $r=\sqrt{R^2+z^2}$ is the intrinsic galactocentric radius (where $R$ is the cylindrical radius and $z$ is the height) and $\rho_{0}$ and $r_{\rm{s}}$ are the central density and the scale radius of the dark matter halo. The last two quantities are inferred using the same procedure explained in \cite{afruni19}, starting from the calculation of the virial mass and radius of the halo. To infer the halo mass we have used, given the properties of our galaxy sample, the stellar to halo mass relation of \cite{posti19a}, obtained through the fit of rotation curves for a sample of low-redshift spiral galaxies with $10^{7}\leq M_{\ast}/M_{\odot}<10^{11}$. We used in particular the linear fit on the same relation performed in \cite{posti19b} (equation B.7). The virial mass is then calculated from $M_{200}$ as in \cite{afruni19}.\\
We acknowledge the simplistic choice of the gravitational potential, which neglects the possible presence of a bulge or other features at the center of our galaxies and employs galaxy discs that are unrealistically thin. This is however justified by the general absence of bulges or bars in our objects (see Figures~\ref{fig:galfit1} - \ref{fig:galfit3} in Appendix~\ref{galfit}) and by the negligible influence that a thicker disc would have on the cloud orbits. The implementation of this simple potential, on the other hand, reduces the computational cost of the integration.
\subsubsection{Integration initial conditions}\label{init}
\begin{figure}
    \includegraphics[width=1\linewidth]{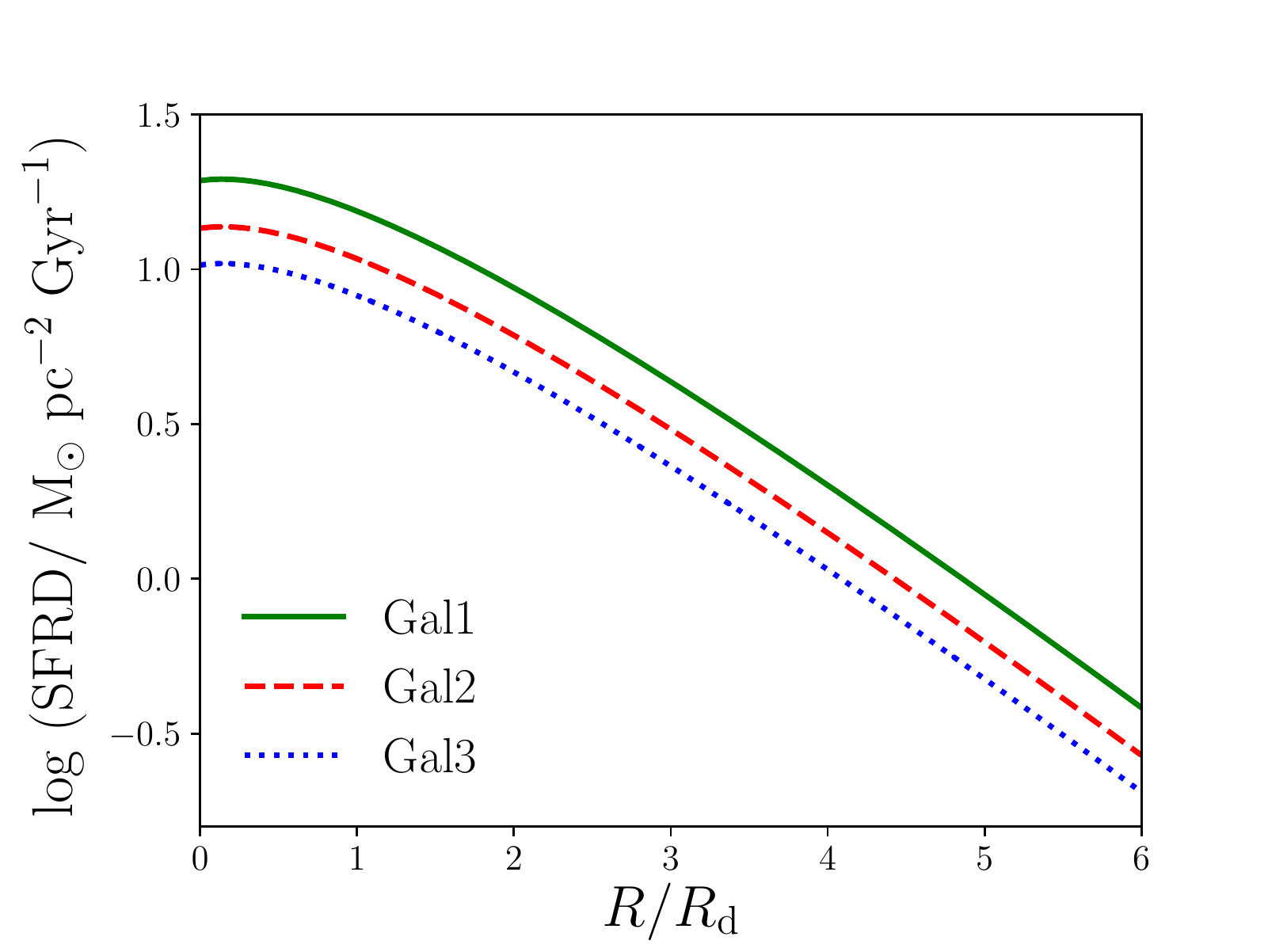}
    \caption{Star formation rate density for the three galaxy models described in Section~\ref{init}, derived following \protect\cite{pezzulli15}.}
              \label{fig:SFRD}%
    \end{figure}

Ideally, one would like to model the cloud orbits for each galaxy, each of them with a different potential, given by the different virial masses and disc radii (see Table~\ref{tab:galprop}
and equations~\ref{eq:potDisc}, \ref{eq:potNFW}). That would however come at a very high computational cost. We therefore made the choice to divide our objects in three subsamples depending on their stellar masses and to create, for each one of these three samples, only one model with a potential calculated using median properties. The same model will be then applied to all the galaxies in the same subsample. We will refer to the three models as Gal1, Gal2 and Gal3 and we list their properties in Table~\ref{tab:modprop}.\\
Once we have defined the potential, which is axisymmetric, the cloud orbits are integrated in the $(R,z)$ (we will refer to these coordinates from here on as $R_{\rm{gal}}$ and $z_{\rm{gal}}$, since they represent the intrinsic reference frame of the galaxy) plane and will be then uniformly distributed across azimuthal angles $\phi$ (see Section~\ref{massrate}). Since the clouds are coming from the supernova explosions from the disc, we assume an initial height $z_{\rm{gal}}=0$, while the initial cylindrical radius $R_{\rm{gal}}$ is randomly selected in the range between 0 and 6 times the disc scale radius $R_{\rm{d}}$, following a probability distribution correspondent to the star formation rate density (SFRD) of each galaxy model\footnote{More in detail, the probability distribution takes into account the geometrical factor and is therefore proportional to $R_{\rm{gal}}$ SFRD.}. This is calculated using the theoretical profile of \cite{pezzulli15}, which was tested on a sample of 35 nearby spiral galaxies. In particular, we used here for each galaxy $\nu_{\rm{M}}=\rm{SFR}/M_{\ast}$ and $\nu_{\rm{R}}=0.35\nu_{\rm{M}}$, where $\nu_{\rm{M}}$ and $\nu_{\rm{R}}$ are respectively the specific mass growth rate and the specific radial growth rate of the disc. This is consistent with a disc inside-out growth, as found by \cite{pezzulli15}. The SFRD profiles of the three galaxy models are shown in Figure~\ref{fig:SFRD}. 
To perform the orbit integration, the initial cloud velocity is needed. We do not assume a fixed value for this velocity, but we let it vary as a free parameter that we call $v_{\rm{kick}}$. Once this value is defined, the three velocity components are obtained through
\begin{ceqn}
\begin{equation}\label{eq:invel}\begin{aligned}
v_{\rm{kick,R}}=v_{\rm{kick}}\sin{\theta}\cos{\phi}\ ,\\
v_{\rm{kick,T}}=v_{\rm{circ}}+v_{\rm{kick}}\sin{\theta}\sin{\phi}\ ,\\
v_{\rm{kick,z}}=v_{\rm{kick}}\cos{\theta}\ ,
\end{aligned}
\end{equation}
\end{ceqn}
where $v_{\rm{circ}}$ is the disc circular velocity\footnote{The circular velocity is assumed positive in our model for all the galaxies, since we do not have information on the direction of the disc rotation. However, using the opposite sign for $v_{\rm{circ}}$, we obtained the same result reported in this paper.}, $\phi$ is randomly selected between 0 and $2\pi$, and $\theta$ is the angle between the direction of the kick and the vertical axis $z_{\rm{gal}}$, ranging between 0 and the angle $\theta_{\rm{max}}$ and randomly selected from a uniform distribution in $\cos{\theta}$. $\theta_{\rm{max}}$ represents the aperture of the outflowing cone of clouds (see Figure~\ref{fig:cone}) and is another free parameter of our analysis.\\
We then created, for each of the three galaxy models that we defined above and for a given choice of $v_{\rm{kick}}$ and $\theta_{\rm{max}}$, $N$ different orbits along this range of initial conditions and we integrate for 10 Gyr. Depending on the initial conditions, the orbits will be either open, meaning that the clouds are escaping the galaxy halos, or closed, with the clouds eventually falling back to the disc. We stop the integration at the moment the clouds reach a distance $r=1.5r_{\rm{vir}}$ from the center or $z_{\rm{gal}}=0$ during their fall.
\subsubsection{Interactions with the hot corona}\label{hydro}
   \begin{figure}
   \includegraphics[width=1\linewidth]{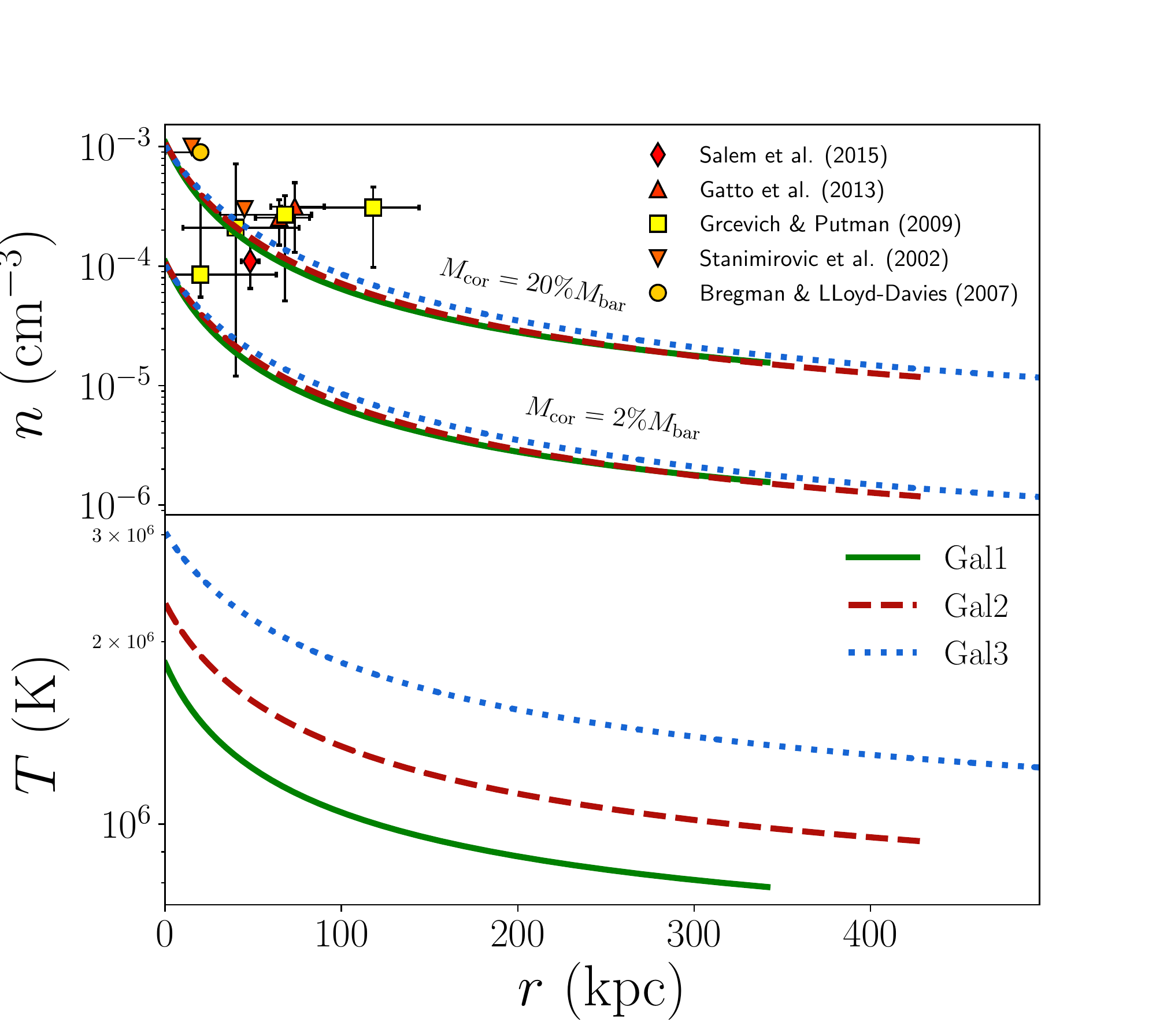}
   \caption{Properties of the hot gas medium for the three galaxy models, respectively density profiles on the top panel and temperature profiles on the bottom panel. The profiles are obtained as described in Section~\ref{hydro}. On the top panel we show both the profiles for a corona bearing 20\% and 2\% of the total baryonic mass expected within the galaxy halo (see main text). The data points represent the observational constraints of \protect\cite{stanimirovic02,bregman07,grcevich09,gatto13,salem15}, all taken from \protect\cite{sormani18}. }
              \label{fig:corona}%
    \end{figure}
\begin{figure}
   \includegraphics[width=1\linewidth]{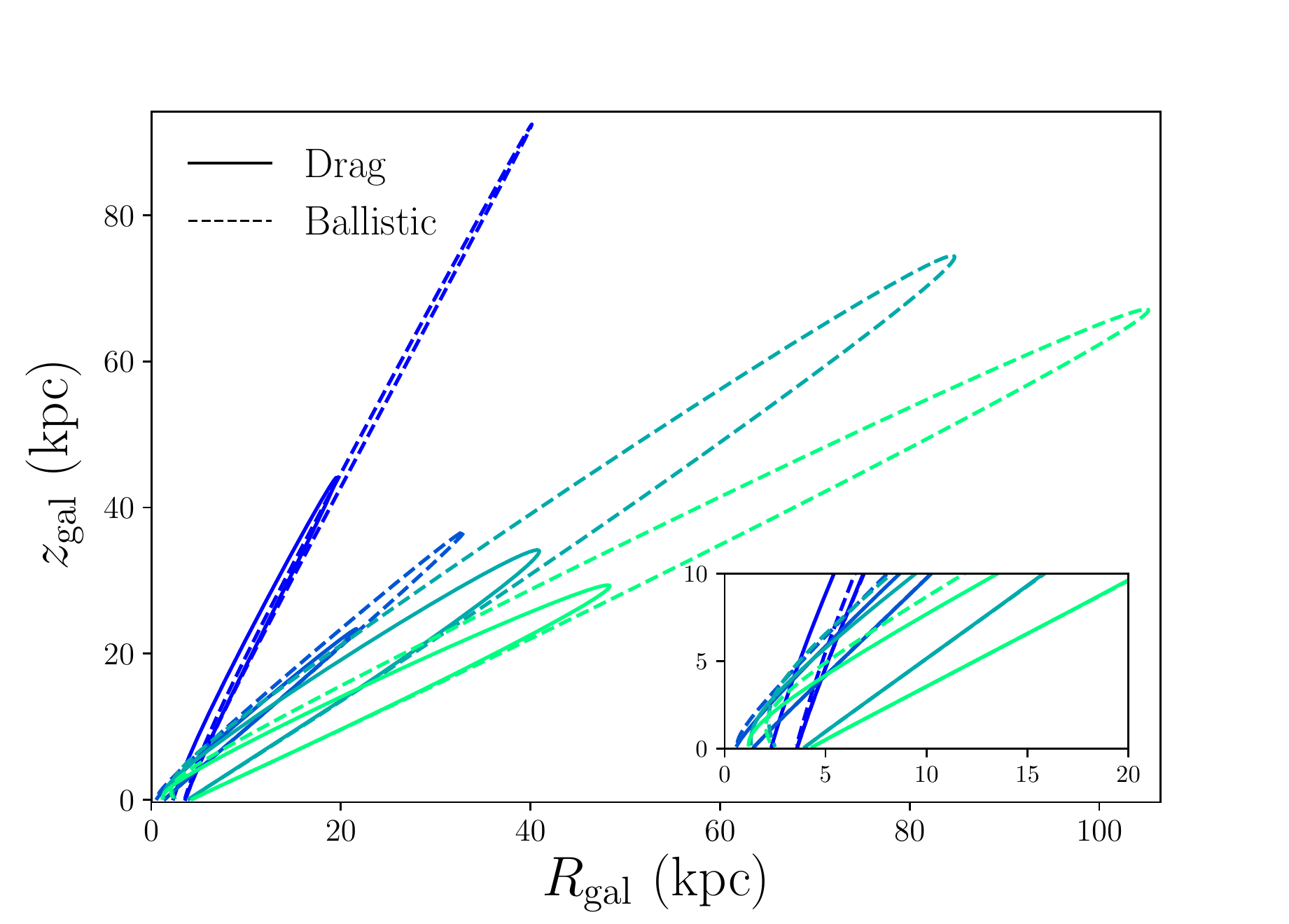}
   \caption{Example of cloud orbits for the model Gal2, with the following choice of parameters: $v_{\rm{kick}}=370\ \rm{km}\ \rm{s}^{-1}$, $\theta_{\rm{max}}=60^{\circ}$ and $m_{\rm{cl}}=10^{6.5}\ M_{\odot}$. The dashed lines represent the prediction of a ballistic model, while the solid ones show the effects of the inclusion of the drag force. The colors represent orbits starting from different positions along the galactic disc and with different angles with respect to the $z_{\rm{gal}}$ axis, selected in the range from 0 to $\theta_{\rm{max}}$. Small panel: zoom-in on the central region of the halo.}
              \label{fig:dragbal}%
    \end{figure}

If gravity (see Section~\ref{gravity}) is the only force driving the motion of the clouds, they would have purely ballistic orbits and the setup described in the previous section would define completely the orbit integration. However, the halos of galaxies are not devoid of gas, but rather filled with a hot medium, the galaxy corona \citep[e.g.][]{anderson11,li17}, at temperatures close to the galaxy virial temperature. In this Section we describe how we model and introduce the coronal gas in our analysis, in order to make the cool absorber dynamics more realistic than the simple ballistic one.\\
We define the corona as a gas in hydrostatic equilibrium with the dark matter halo described by equation \eqref{eq:potNFW}. More in detail, the hot gas density profile is described by \citep{binney09}\\
\begin{ceqn}
\begin{equation}\label{eq:corn}
\frac{n_{\rm{e}}(r)}{n_{\rm{e},0}}=\left(\frac{T}{T_0} \right)^{1/(\gamma-1)}\ ,
\end{equation}
\end{ceqn}
where
\begin{ceqn}
\begin{equation}\label{eq:corT}
\frac{T(r)}{T_0}=\frac{\gamma-1}{\gamma}\frac{\mu m_{\rm{p}}}{k_{\rm{B}} T_0}\left(\Phi(r)-\Phi_0 \right)\ .
\end{equation}
\end{ceqn}\\
Here, $\Phi(r)$ is the NFW potential, $m_{\rm{p}}$ is the proton mass, $\mu=0.6$ is the mean molecular weight, $\gamma$ is the polytropic index and $T_0$, $n_{\rm{e},0}$ and $\Phi_0$ are respectively the temperature, the density and the potential at the reference radius $r_0=10\ \rm{kpc}$. The polytropic index and the two normalization factors are chosen in order to have temperature and density profiles consistent with the (uncertain) observational constraints (see Figure~\ref{fig:corona}, where the observational data points are taken from \citealt{sormani18}). More in detail, we use $\gamma=1.2$, which allows the coronal temperature to vary throughout the halo, without implying too large variations in the density and in the temperature profiles between the internal and the external regions, as shown in the two panels of Figure~\ref{fig:corona}. With this choice, the density of the inner regions is consistent with the observational values and it slightly decreases with the galactocentric radius.
An isothermal corona ($\gamma=1$) at a temperature close to the virial one would have central densities too high to be reconciled with the observations \citep[e.g.][]{salem15}, while using a higher polytropic index would lead to unrealistically low densities in the external regions of the halos. The density normalization is set in order to have a total mass of the hot gas equal to 20\% of the baryonic mass within the galaxy halo, which is a fraction $f_{\rm{bar}}$ of the galaxy virial mass, where $f_{\rm{bar}}=0.158$ is the cosmological baryon fraction \citep{planck18}. This choice leads to values of the density that are compatible with the observations, as can be seen in Figure~\ref{fig:corona}. We will relax this assumption in Section~\ref{diffcor}. Regarding the temperature of the hot gas, we set an inner temperature $T_0=2.8T_{\rm{vir}}$. The model temperature slightly decreases with the distance from the central galaxy, remaining close to the galaxy virial temperature, as from theoretical expectations \citep[e.g.][]{white78,fukugita06}.\\
Once the density and the temperature of the corona are defined, the density of the cool CGM is obtained by imposing pressure equilibrium between the hot gas and the cool clouds, assumed to be at a temperature of $2\times 10^4$ K, in agreement with observational estimates \citep{werk13,keeney17,lehner18}. The main effect of the hot gas on the clouds is to slow them down by means of the drag force, given by \citep[see][]{marinacci11,afruni19}
\begin{ceqn}
\begin{equation}\label{eq:dragforce}
\dot{v}_{\rm{drag}}=-\frac{\pi r^2_{\rm{cl}}\rho_{\rm{cor}}v^2}{m_{\rm{cl}}}\ ,
\end{equation}
\end{ceqn}
where $v$ is the relative velocity between the clouds and the corona, $m_{\rm{cl}}$ is the cloud mass, $r_{\rm{cl}}$ is the cloud radius (set by the choice of the mass and the pressure equilibrium, see \citealt{afruni19}) and $\rho_{\rm{cor}}=\mu m_{\rm{p}}n_{\rm{cor}}$ is the hot gas mass density, with $\mu=0.6$ and $n_{\rm{cor}}=2.1n_{\rm{e}}$. More massive clouds will be less affected by the interactions with the hot corona, while the motion of less massive clouds will be strongly influenced by the ambient gas. In general, with respect to the ballistic case, the clouds will need higher kick velocities to reach the external parts of the halos. The mass of the clouds $m_{\rm{cl}}$ is the third free parameter of our models. We implemented in GALPY an additional part of the equation of motion of the cool clouds that takes into account the drag force acted by the hot coronal gas whose properties are described by equations~\eqref{eq:corn} and \eqref{eq:corT}, to obtain more realistic results from the orbit integration.\\
In Figure~\ref{fig:dragbal} we show the influence of the drag force on the cloud orbits, for the model Gal2. We observe a similar behavior for the other two models. 
As a reference we chose, to create these orbits, $v_{\rm{kick}}=370\ \rm{km}\ \rm{s}^{-1}$, $\theta_{\rm{max}}=60^{\circ}$ and $m_{\rm{cl}}=10^{6.5}\ M_{\odot}$. The dashed curves represent the ballistic orbits, while the solid curves show the results of the scenario where the corona is affecting the motion of the clouds. In each of the two scenarios, the different distances from the central galaxy reached by the orbits are mostly due to the initial cylindrical radius $R_{\rm{gal}}$ from which the clouds are ejected: orbits starting from larger radii will reach larger distances, due to the lower pull of the gravitational potential. All the orbits ends back to the central regions of the disc, with no substantial difference between the drag and ballistic scenarios, as can be seen in the zoom-in panel in the lower right part of Figure~\ref{fig:dragbal}. 
The main difference between the two models is that in the case including the drag force of the hot gas the clouds reach much smaller distances from the central galaxy. To reach the distances that we see in the observations (up to the galaxy virial radii) we will therefore need kick velocities significantly higher than what we would expect from a purely ballistic model.\\
In reality, the drag force is not the only effect that the corona has on the cloud motion. In fact, our modeling does not take into account all the hydrodynamic instabilities that take place at the interface between the two gas phases \citep[see][]{armillotta17,gronnow18,gronke18}, as well as other effects like the thermal conduction. A full hydrodynamic treatment is outside the scope of this work, given the complications and uncertainties that it would imply. As we will discuss in Section~\ref{dischydro}, with a more rigorous treatment of the hydrodynamics the clouds would likely need even higher velocities to be ejected out to the same distances. 
\subsection{Outflow rate}\label{massrate}
\begin{figure}
   \includegraphics[width=1\linewidth]{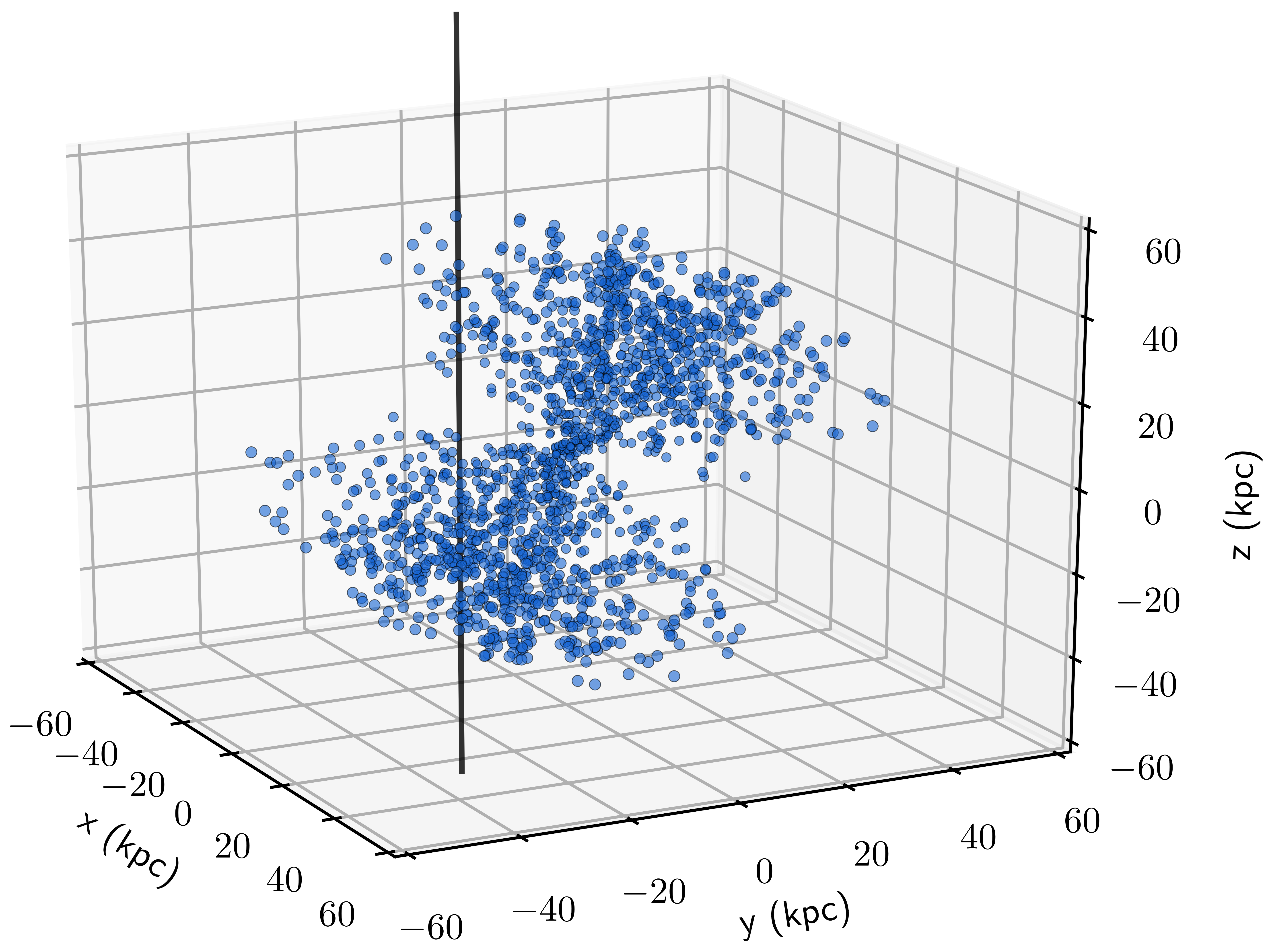}
   \caption{Cloud population for the same model used to create Figure~\ref{fig:dragbal}, with $\eta=2$ and a disc inclination $i=30^{\circ}$. The clouds are outflowing from the galaxy in a biconical shape. The black line represents one of the lines of sight that we used to perform our synthetic observations.}
              \label{fig:cone}%
    \end{figure}

Once the orbits have been calculated, we need to populate them with clouds and this requires the knowledge of the mass outflow rate from the galaxy. We implement a rate of mass ejection from the disc that is constant with time. Since we assume that the cool CGM comes from supernova feedback, we relate the mass outflow rates to the star formation rates of the central galaxies reported in Table~\ref{tab:modprop}, through the formula\\
\begin{ceqn}
\begin{equation}\label{eq:massrate}
\dot{M}_{\rm{out}}=\eta\  \rm{SFR}\ ,
\end{equation}
\end{ceqn}\\
where $\eta$ is the mass loading factor and is the fourth and last parameter of our models. Dividing half of the mass outflow rate for the mass of the clouds $m_{\rm{cl}}$ we obtain the total number of clouds $\dot{n}_{\rm{out}}$ ejected from one side of the disc per unit of time and we assume that these clouds are uniformly distributed with respect to time along the $N$ orbits that we are modelling. As explained in Section~\ref{init}, the integration of each orbit is stopped once the clouds have fallen back to the galactic disc or reached 1.5 times the galaxy virial radius: the integration time $t_{\rm{orb}}$ will then be different for every orbit. The number of clouds for each orbit is therefore given by\\
\begin{ceqn}
\begin{equation}
n_{\rm{orb}}=\frac{\dot{n}_{\rm{out}}}{N}t_{\rm{orb}}\ . 
\end{equation}
\end{ceqn}\\
Each of these clouds is placed in the orbit at a different time, separated by $\Delta t=t_{\rm{orb}}/n_{\rm{orb}}$, and has the properties (position in the $R_{\rm{gal}} - z_{\rm{gal}}$ plane, velocity components, density, radius) predicted by our model at that time. To each of the clouds is then assigned a random azimuthal position $\phi$ ranging from 0 to $2\pi$ and the same procedure is performed for both sides of the disc \citep[see][]{fraternali06,fraternali08}. Throughout this work, we use for the number of orbits $N=30$, but our results do not depend on the choice of this number.\\
We show in Figure~\ref{fig:cone} the result in 3D of the treatment explained above, for the same choice of parameters as in Figure~\ref{fig:dragbal} and with $\eta=2$: the clouds are distributed in a cone-like structure on both sides of the galactic disc. Note that
the intrinsic reference frame of the galaxy can be different from the frame $(x,y,z)$ of the observations, depending on the galaxy inclination (see Table~\ref{tab:galprop}).
   \begin{figure*}
   \includegraphics[width=1\linewidth]{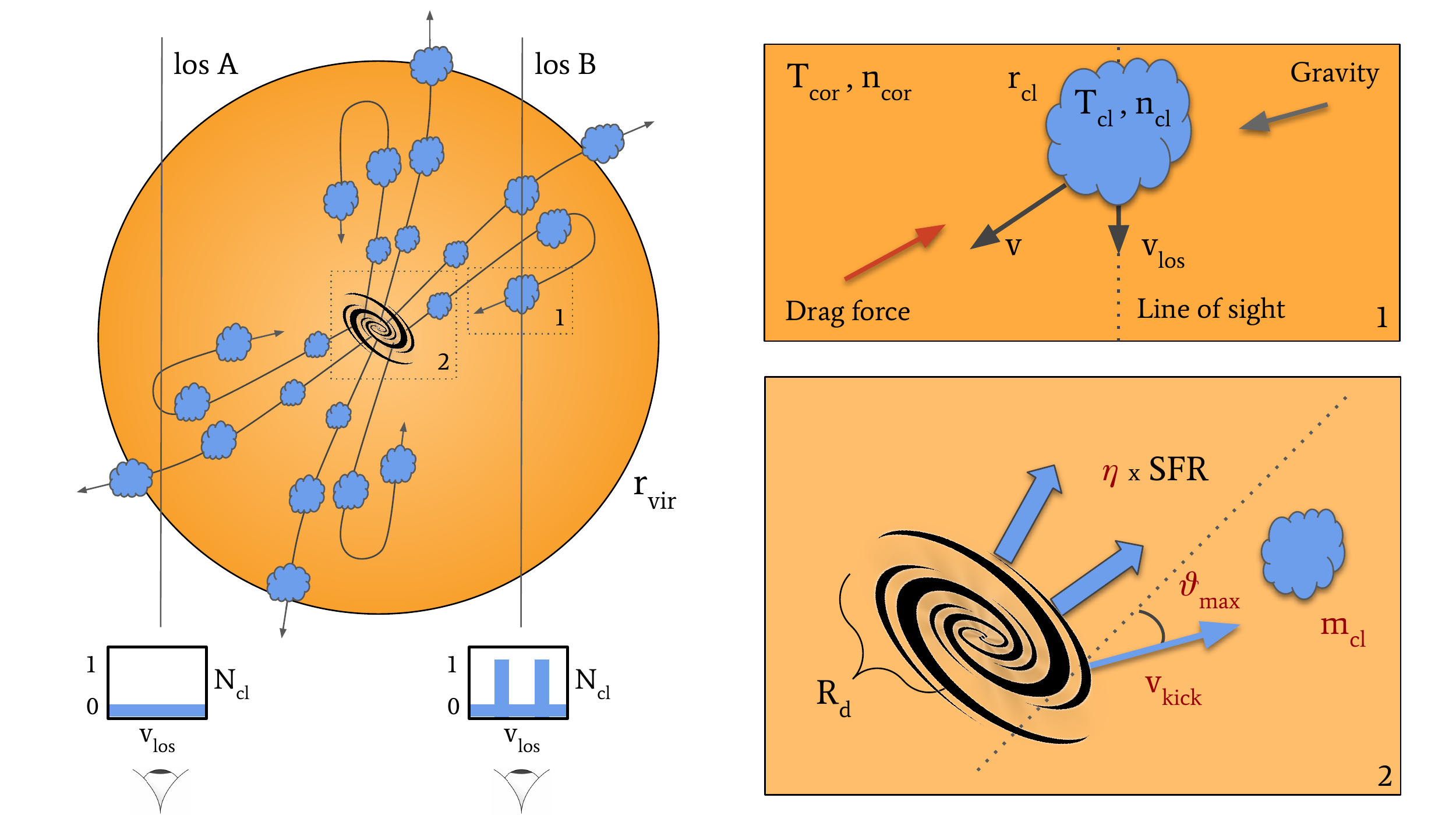}
   \caption{Diagram summarizing the modeling used in this work and described in Section~\ref{model}. Left diagram: representation of the biconical outflow of clouds ejected from the central galaxy, with an example of synthetic observations along two different lines of sight. Line of sight (los) A does not intercept any cloud and therefore the resultant velocity distribution is empty, while line of sight (los) B intercepts two clouds, resulting in two different velocity components. Top right panel: zoom-in on a single cool circumgalactic cloud, with temperature $T_{\rm{cl}}=2\times10^4\ \rm{K}$ and density and radius given by the pressure confinement of the hot coronal gas, whose density and temperature are found respectively through equations~\eqref{eq:corn} and \eqref{eq:corT}. The cloud is pulled towards the disc by the gravitational force, while the hot gas, through the drag force (equation~\ref{eq:dragforce}), slows it down along its entire orbit. The velocity $v_{\rm{los}}$, which will be compared with the real data, is the component along the line of sight of the total cloud velocity. Bottom right panel: zoom-in on the central galaxy. The four free parameters of our models ($m_{\rm{cl}}$, $v_{\rm{kick}}$,  $\eta$, $\theta_{\rm{max}}$) are depicted in red. The total cold mass outflow rate is proportional to the galaxy star formation rate and the starting points of the orbits are distributed along the disc following the star formation rate density of \protect\cite{pezzulli15}.}
              \label{fig:diagram}%
    \end{figure*}
\subsection{Comparison with the observations}\label{compar}
The idea behind our analysis is that our models depend on parameters that define different physical scenarios and that we let free to vary. Through the comparison of our model outputs with the COS observations, we can find the best choice of parameters and therefore the dynamical scenario that better describes the observed kinematics of the cool CGM around star-forming galaxies. In this Section we explain how we perform this comparison.
\subsubsection{Synthetic observations}\label{synthetic}
To compare our results with the observations we performed synthetic observations, using the cloud populations created as explained in Section~\ref{massrate}. As already mentioned, these are obtained in the reference frames of the galaxies, which are different from the one of the observations. The lines of sight intersect a plane $(x,y)$ that coincides with the plane $(x_{\rm{gal}},y_{\rm{gal}})$, with $x_{\rm{gal}}=R_{\rm{gal}}\cos{\phi}$ and $y_{\rm{gal}}=R_{\rm{gal}}\sin{\phi}$, only if the galaxy is face on, with $z=z_{\rm{gal}}$. This is however not the case for most of our galaxies, as found with the GALFIT analysis (see Appendix~\ref{galfit} and Table~\ref{tab:galprop}), and we can see in Figure~\ref{fig:cone} how the direction of the outflowing cones does not match the direction of the line of sight (the inclination of the disc used to create this figure is equal to $30^{\circ}$). The first step to perform the observations of our model halos is to transform the reference frame of the galaxies into the one of the observations, through
   \begin{figure*}
   \centering
   \includegraphics[clip, trim={0 0 0cm 0}, width=13.3cm]{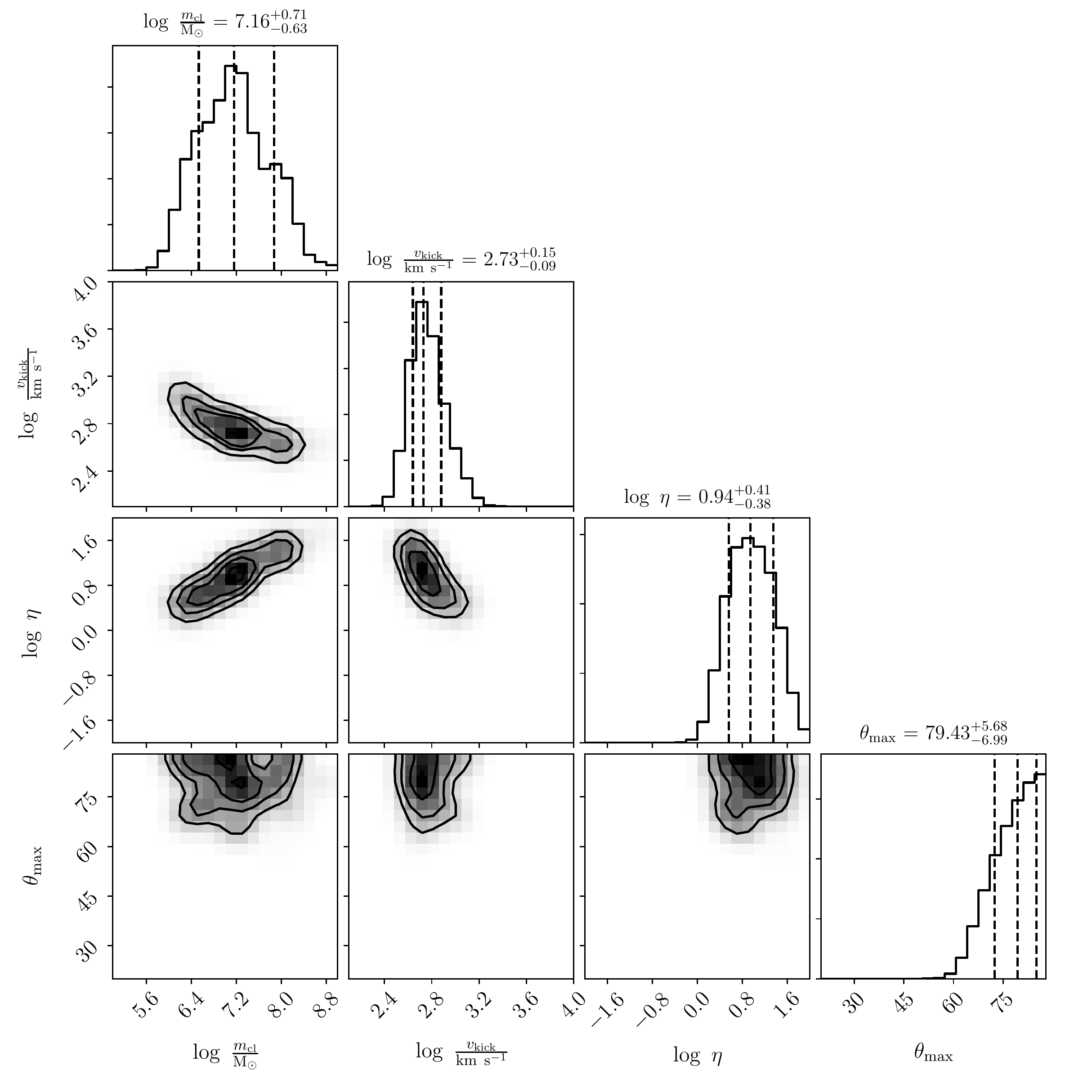}
   \caption{Corner plot with the MCMC results, representing the one and two dimensional projections of the posterior probabilities for the four free parameters of our models. The parameter space is explored in the logarithm of the angle $\theta_{\rm{max}}$, but the results are transformed here in physical units for clarity.}
              \label{fig:MCMC}%
    \end{figure*}
\begin{ceqn}
\begin{equation}
y=y_{\rm{gal}}\cos{i}+z_{\rm{gal}}\sin{i}\ ,
\end{equation}
\end{ceqn}
while we set $x=x_{\rm{gal}}$. This transformation is applied to all the galaxies in Table~\ref{tab:galprop}.
Once we derived the position $(x,y)$ of each cloud, we then traced line of sights at the same positions of the observations and we picked all the clouds intercepted (the distance of the position of the cloud from the position of the line of sight is less than the radius of the cloud) and their line-of-sight velocity, through the formula
\begin{ceqn}
\begin{equation}
v_{\rm{los}}=-v_{\rm{y,gal}}\sin{i}+v_{\rm{z,gal}}\cos{i}\ ,
\end{equation}
\end{ceqn}\\
where $v_{\rm{y,gal}}=v_{R}\sin{\phi}+v_{T}\cos{\phi}$ and $v_{\rm{z,gal}}=v_{\rm{z}}$. With this treatment, we end up having for each line of sight the kinematic prediction of our model, directly comparable with the observations outlined in Section~\ref{Observations}. The creation of the synthetic observations, along with a visualization of the main parameters and properties of our models, is summarized in the diagram of Figure~\ref{fig:diagram}.
\subsubsection{Likelihood}\label{likelihood}
To find the best model that reproduces the observations shown in Section~\ref{Observations} we performed a bayesian Markov Chain Monte Carlo (MCMC) analysis over the 4-dimensional space defined by the four free parameters of our modeling: $m_{\rm{cl}}$, $v_{\rm{kick}}$, $\eta$ and $\theta_{\rm{max}}$. In order to achieve this, we compared the results obtained with the synthetic observations outlined in Section~\ref{synthetic} with the actual COS data. The comparison was done through a likelihood that takes into account the number and velocity distribution of the absorbers observed along each individual line of sight. In particular we developed a technique to compare for each sightline the predictions for the CGM of our modeled galaxies with the observations. We call the likelihood of a single sightline $\mathcal{L}_{\rm{los}}$. The total likelihood that we will use for the bayesian analysis is given by the product of all the 41 single likelihoods, in order to deploy all the kinematic constraints coming from different projected distances from the central galaxies.\\
The likelihood $\mathcal{L}_{\rm{los}}$ can be divided into the products of two different terms, that we call $\mathcal{L}_{\rm{num}}$ and $\mathcal{L}_{\rm{kin}}$, which respectively represent the comparisons between the numbers of components and the kinematic distributions of model and observations.
More in detail, for each line of sight we created a velocity distribution over the range from -600 to 600 $\rm{km}\ \rm{s}^{-1}$, using the line-of-sight velocities calculated as explained in Section~\ref{synthetic}. We divided this range in 24 bins, in order to have a bin width of 50 $\rm{km}\ \rm{s}^{-1}$, which is consistent with the average line width of the observed absorption lines: $53\ \rm{km}\ \rm{s}^{-1}$ for the COS-GASS sample \citep[from][]{borthakur16} and $30\ \rm{km}\ \rm{s}^{-1}$ for the COS-Halos sample \citep[from][]{tumlinson13}. In particular, in order to have statistically significant distributions, for each choice of parameters we averaged the outputs of 50 different realizations of the same model\footnote{With 50 realization we are able to take into account the fluctuations of the model, as proven by the successful tests carried out in Appendix~\ref{artdata}.}, since one individual model can be affected by fluctuations due to the intrinsic randomness of the cloud positions along the orbits (see Section~\ref{massrate}).\\ 
For the first term of the likelihood, we used the Poisson statistics to compare the number of observed ($n_{\rm{obs}}$, see Table~\ref{tab:galprop}) and model components $n_{\rm{mod}}$, the latter given by the number of bins of the model velocity distribution with at least one cloud (in particular, $n_{\rm{mod}}$ is the mean value of the 50 model realizations that we are using for the comparison). This comparison is therefore given by
\begin{ceqn}
\begin{equation}\label{eq:likenumb}
\mathcal{L}_{\rm{num}}=n_{\rm{mod}}^{n_{\rm{obs}}}\ \frac{e^{-n_{\rm{mod}}}}{n_{\rm{obs}}!}\ ,
\end{equation}
\end{ceqn}
where $n_{\rm{obs}} !$ is the factorial of the observed number of components. Since $\mathcal{L}_{\rm{num}}$ is not defined for $n_{\rm{mod}}=0$ and $n_{\rm{obs}}\neq 0$, in these cases we defined $n_{\rm{mod}}=1/50$, where 50 is the number of realizations.
The second term is instead given by the bayesian probability of the observed velocity components given our model. In particular the probability for each component is given by the value of the normalized model velocity distribution in the bin where that velocity is observed. We can then obtain the value of $\mathcal{L}_{\rm{kin}}$ through the product of the $n_{\rm{obs}}$ probabilities predicted by our model for each line of sight.\\
Once the two terms are defined, the total likelihood of the single line of sight is obtained through
\begin{ceqn}
\begin{equation}\label{eq:lnliketot}
\ln\mathcal{L}_{\rm{los}}=(\ln\mathcal{L}_{\rm{num}}+\ln\mathcal{L}_{\rm{kin}})/(1+n_{\rm{obs}}) ,
\end{equation}
\end{ceqn}
where the weight in the denominator takes into account for the number of constraints on each line of sight.\\
It is important to mention that, for each line of sight, the prediction of the model depends on the sign of the inclination $i$ with respect to the plane of the sky. From the GALFIT analysis, we are not able to disentangle what is the sign of the inclination and therefore the direction of the outflow cones (in Figure~\ref{fig:cone}, the direction of the two cones would be symmetric to the current one with respect to the z-axis if we chose an opposite sign for the inclination). For each line of sight, we therefore performed our synthetic observations for both inclinations and we kept the one with the highest $\mathcal{L}_{\rm{los}}$, i.e. the one more similar to the observations.\\
We have tested the likelihood explained above on a number of artificial data sets created with our models, in order to verify whether with this analysis we are able to properly constrain the 4 free parameters of our model. We found that the initial set of parameters can be successfully recovered by our MCMC analysis using the likelihood outlined above. The results of these tests are presented in detail in Appendix~\ref{artdata}. In the next Section we will show instead what we find when we apply this likelihood to the data shown in Section~\ref{Observations}.
\section{Results}\label{results}
   \begin{figure*}
   \includegraphics[clip, trim={0.2cm 0 0cm 0.2cm}, width=0.49\linewidth]{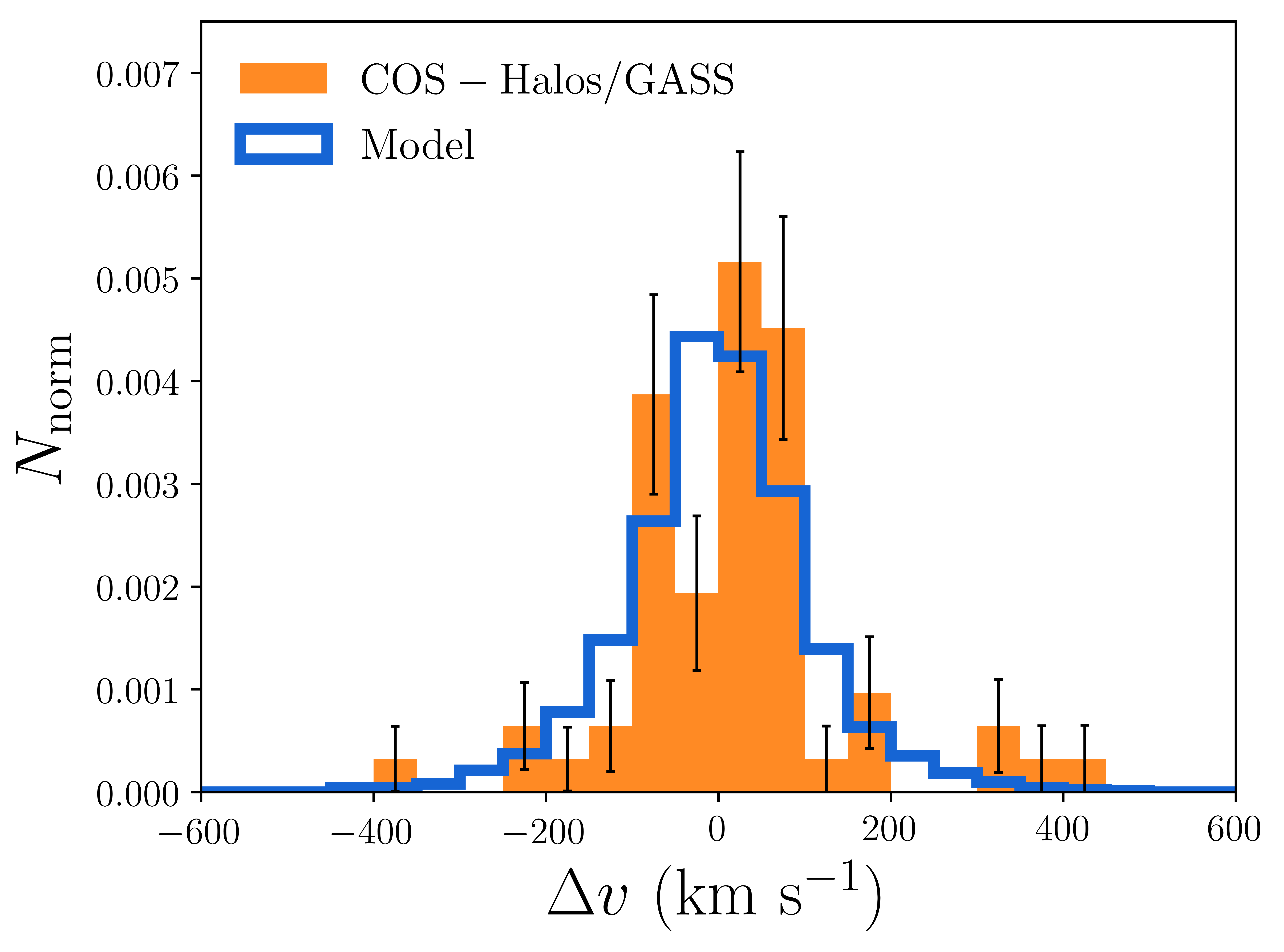}
   \includegraphics[clip, trim={0cm 0.1cm 0cm 0cm},width=0.48\linewidth]{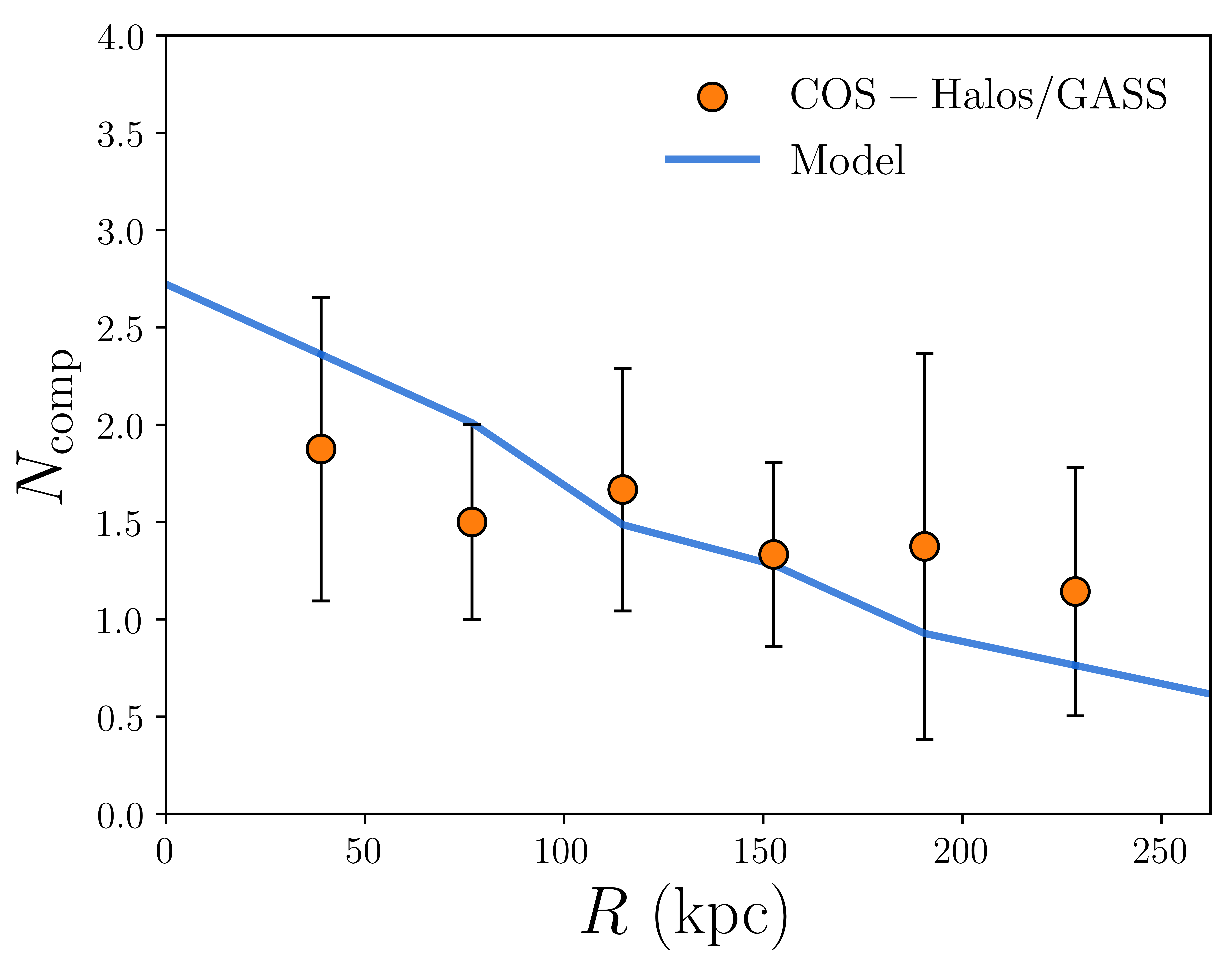}
   \caption{Comparison between the outputs of the best models found with the MCMC analysis (blue) and the observations (orange), see Section~\ref{mcmc} for more details. Left: total velocity distributions, with the errors in the observations calculated through bootstrapping. Right: average number of components per line of sight in radial bins of 35 kpc, as a function of the projected distance from the central galaxy (the blue line connects the model values predicted for each bin). The uncertainties are given by the standard deviation of the number of observed components for each bin.}
              \label{fig:PV}%
    \end{figure*}
In this Section we report the results of the MCMC analysis on the COS data that we have performed over the parameter space using the likelihood defined in Section~\ref{likelihood} and we discuss the physical meaning of the scenario described by the models that best reproduce the observations.
\subsection{MCMC analysis}\label{mcmc}
\begin{figure}
   \includegraphics[clip,trim={0 0 0cm 2cm},width=1\linewidth]{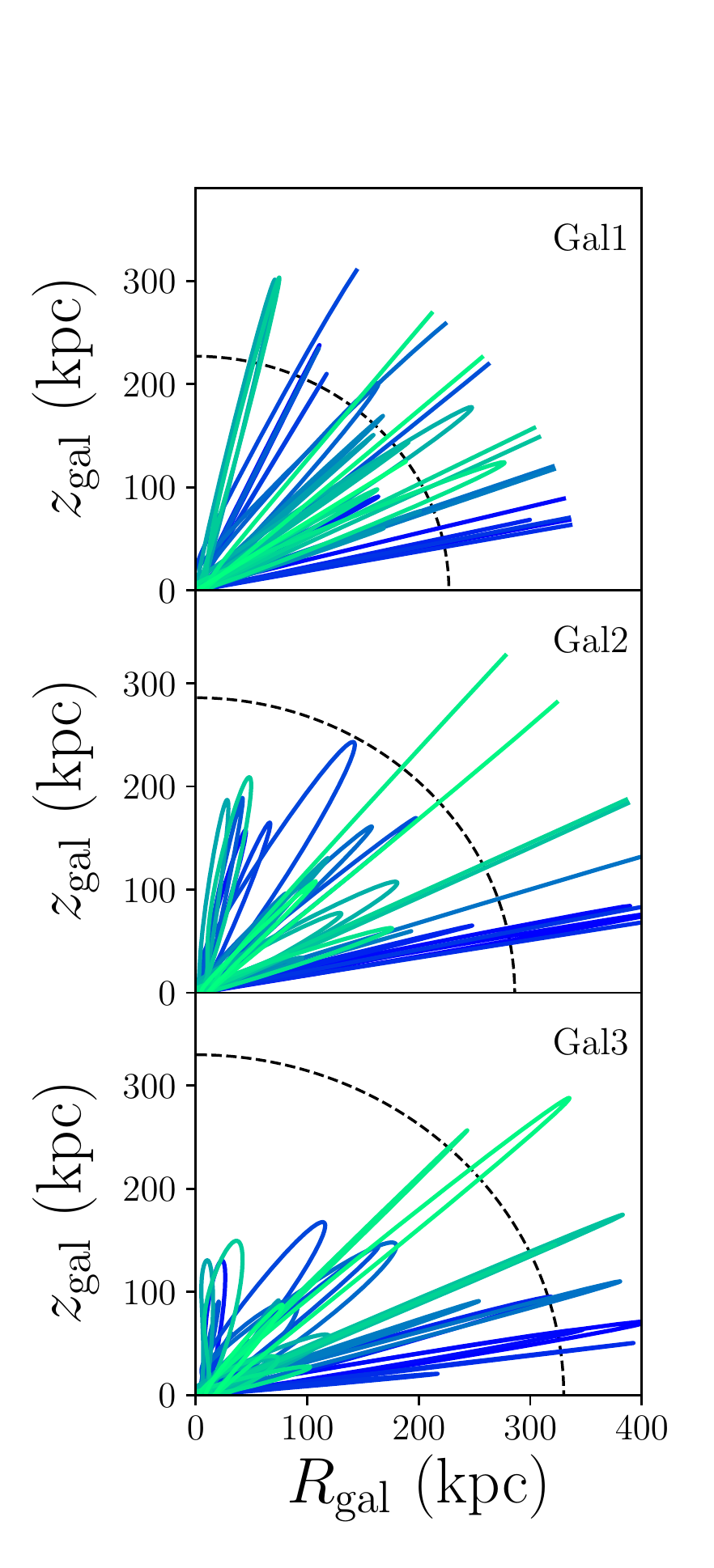}
   \caption{Representative orbits of the clouds for the best models, obtained using the median value of the 4 paramaters reported in Figure~\ref{fig:MCMC}. The three panels show the results of Gal1,Gal2 and Gal3 and the dashed curves show the value of the virial radius for each of the three galaxies. As in Figure~\ref{fig:dragbal}, different orbits show different colors.}
              \label{fig:res_orbit}%
    \end{figure}

We explored the 4-dimensional parameter space over the following ranges:
\begin{itemize}
    \item $5<\log(m_{\rm{cl}}/M_{\odot})<9$ ,
\vspace{1mm}
    \item $2<\log(v_{\rm{kick}}/(\rm{km}\ \rm{s}^{-1}))<4$ ,
\vspace{1mm}
    \item $-2<\log \eta<2$ ,
\vspace{1mm}
    \item $\log 20^{\circ}<\log \theta_{\rm{max}}<\log 90^{\circ}$ ,
\end{itemize}
using flat priors for all the parameters in the logarithmic space. In Figure~\ref{fig:MCMC} we report the one and two dimensional projections of the posterior distributions for the four parameters, with the values of the 32th, 50th and 68th percentiles (also reported in Table~\ref{tab:testres}).
Note from Figure~\ref{fig:MCMC} that there is a very well defined region of the parameter space where the posterior is maximized:
the models with this choice of parameters represent the physical scenario that best reproduces the observations.\\
In Figure~\ref{fig:PV} we show how the results of our best models compare with the observational data that we have used in this work, displaying in particular on the left the total velocity distribution of the cool gas absorbers and on the right the number of components as a function of the projected distance from the central galaxy. The results are averaged over 100 different models with the 4 free parameters ranging in the area of the parameter space highlighted in Figure~\ref{fig:MCMC} within the 32nd and 68th percentiles of the posterior distributions.
The observed velocity distribution, in orange, is the same as the one shown in Figure~\ref{fig:velobs}, while the model distribution is obtained combining all the velocities obtained for each line of sight using the technique explained in Section~\ref{synthetic}. We can note how with our analysis we have found models for which the total kinematic distribution of the cool gas clouds is consistent with the observations. A Kolmogorov-Smirnov test confirmed that the two distributions of observations and model are consistent with each other, with a probability value $p=0.25$. To obtain the plot in the right panel of Figure~\ref{fig:PV} we divided the radial range into uniform bins of 35 kpc, each of them containing a certain number of sightlines. Both for model and observations we calculated the average number of components per line of sight in each bin, with the uncertainties given by the standard deviation of the observations. We can see how, also in this case, the model predictions (blue line connecting the average model values of each bin) are consistent with the observations (orange points), with the number of components decreasing as the distance from the galaxy increases.
We therefore conclude that an outflow scenario, using a very particular choice of physical parameters, is overall able to reproduce the observational features of the COS data of the cool CGM around star-forming galaxies.\\
In Figure~\ref{fig:res_orbit} we can look more in detail at the properties of the models outlined above, in particular using the median value of the posterior probabilities of the 4 parameters.
One peculiarity of the observations outlined in Section~\ref{Observations} is that the cool absorbers are observed till very large distances from the central galaxies (see Figure~\ref{fig:Obslosplane}).
Therefore, to be able to reproduce these data, the orbits derived with our models should reach these large distances: this is visible in the three panels of Figure~\ref{fig:res_orbit}, where we show the orbits described by the clouds for the three galaxy models. The higher the mass of the galaxy, hence the virial mass of the halo, the stronger is the gravitational pull and the harder is for the clouds to travel till distances comparable to the virial radius. Moreover, the COS data present only one non-detection, with the cool CGM observed ubiquitously over the $(x,y)$ plane (Figure~\ref{fig:Obslosplane}). In order to match this feature of the observations, our models require very large apertures for the outflow cones, with $\theta_{\rm{max}}\gtrsim80^{\circ}$. The outflows are therefore not collimated along the minor axis of the galaxies and they are instead more isotropically distributed. The different length of orbits in the same galaxy model is mainly due to the different initial $R_{\rm{gal}}$ of each orbit, since the clouds are distributed along the disc in the range going from 0 to 6 times the disc scale radius, as explained in Section~\ref{init}. More external orbits will experience a weaker gravitational pull and therefore, at equal ejection velocity, will travel to larger distances.\\
We can see from Figure~\ref{fig:res_orbit} how some of the orbits are open, with the clouds escaping outside the virial radius (represented by the dashed curve in each panel) and never coming back to the central galaxy, while other orbits describe a cycle in which the clouds are travelling to very large distances and eventually fall back towards the galactic disc. To reach these distances the cloud need to have large masses, in order to minimise the drag force (see equation~\ref{eq:dragforce}) acted by the corona, and initial velocities of more than $500\ \rm{km}\ \rm{s}^{-1}$. We will see in the next section the physical implications of these values of the parameters.\\ 
\subsection{Physics of the outflows}\label{physout}
In Section~\ref{mcmc} we have seen that outflow models of cool clouds can reproduce the COS-Halos and COS-GASS kinematic data. In this section we look instead at the implications that this model would have for the efficiency of star formation feedback. The value of the 4 parameters has very important implications from an energetic point of view. The kinetic energy produced by supernovae per unit time and available for the wind is given by \citep{cimatti19}
\begin{ceqn}
\begin{equation}\label{eq:snpower}
\dot{K}\approx 3\times10^{40} \left(\frac{f_{\rm{SN}}}{0.1 }\right) \left(\frac{E_{\rm{SN}}}{10^{51}\ \rm{erg}}\right)\left(\frac{\rm{SFR}}{M_{\odot}\ \rm{yr}^{-1}} \right)\ \rm{erg}\ \rm{s}^{-1} ,
\end{equation}
\end{ceqn}
where $f_{\rm{SN}}$ is the efficiency of the supernovae in transferring energy to the wind and $E_{\rm{SN}}$ is the amount of energy released by one supernova explosion. We can estimate the efficiency predicted by our models by calculating the kinetic power of the outflowing wind, which can be expressed as
\begin{ceqn}
\begin{equation}\label{eq:outpower}
\dot{K}_{\rm{out}}= \frac{1}{2} \dot{M}_{\rm{out}}\ v_{\rm{kick}}^2\ ,
\end{equation}
\end{ceqn}
where $\dot{M}_{\rm{out}}$ is the mass outflow rate as defined in equation~\eqref{eq:massrate}. The efficiency necessary to reproduce the cool CGM clouds with our outflow models will then be given by the ratio between equations~\eqref{eq:snpower} and \eqref{eq:outpower}. Using as a kick velocity and as a mass loading factor the best values found with the MCMC analysis and the canonical value $E_{\rm{SN}}=10^{51}\ \rm{erg}$, we obtain $f_{\rm{SN}}\sim2.5$, which corresponds to an efficiency of energy transfer from the supernova explosions to the gas wind of about $250\%$. Clearly, such a value is not physically justifiable, since it means that the outflows would need more energy than the one available from the supernovae. Moreover, from a theoretical point of view the supernovae are expected to radiate away most of their energy \citep[e.g.][]{mckee77} and roughly only 10\% is expected to be transferred to the gas as kinetic energy \citep{kim15, martizzi16,bacchini20}. Recent simulations \citep{fielding18} show that this number could increase to 20-30\% if we consider spatial and temporal clustered supernovae explosions, but even these enhanced efficiencies are still far lower than the one that would be needed to reproduce the observations in an outflow scenario, as we have found. Other stellar feedback modes, like winds from massive stars, can certainly not account for this discrepancy, since the SNe are by far the dominant source of energy over the other mechanisms \citep{elmegreen04}. The efficiency would be significantly reduced if the circumgalactic clouds were originated in a period in which the star formation rates of the galaxies were much higher than the current ones. This would reduce the mass loading factors needed to reproduce the observations. However, the typical timescales that our models predict for the clouds to travel from the central galaxy to their current positions are of the order of two Gyr or less and we do not expect all the galaxies in our sample to have had significantly different star formation rates during this period. Even though sporadic fast outbursts of star formation could have happened in the last two Gyr, it is highly unlikely that they can explain the entirety of the cool circumgalactic gas, that is observed almost ubiquitously at any line of sight intersecting the galaxy halos.\\
We have found therefore that even though the outflow models can reproduce the cool CGM observations, the implications of these models are unfeasible from an energetic point of view. In particular, the velocities and mass loading factors required to reproduce the data lead to unphysical scenarios that need more energy than the one available from the stellar feedback of the central galaxy. Moreover, the scenario described in Section~\ref{mcmc} is unlikely to be a realistic representation of the CGM also because of other properties of the outflowing clouds. In particular, we have seen that to reach the distances seen in the observations, the clouds need to be extremely massive in order to overcome the deceleration acted by the coronal drag force (see Section~\ref{hydro}). With our fitting analysis we obtain a cloud mass of about $10^7\ \rm{M}_{\odot}$, a value that is much higher than the typical masses (which are however very uncertain) expected for these clouds \citep[see][]{werk14, keeney17}. These very high masses lead the clouds to have radii that go from about 2 kpc in the inner regions to 7 kpc in the external parts, where the densities are lower. Even though there is some observational evidence \citep{rubin18} of similar scales for the cool CGM absorbers, it is unlikely that the majority of the clouds have such large radii, in particular the ones just ejected out of the galaxy. Each of these clouds would have a size comparable with the one of the region of the disc from which they were all produced in the first place. The picture described by these models is therefore unrealistic and hardly justifiable.\\
We conclude that an outflow of clouds driven by star formation in the galactic disc is not a realistic scenario to describe the dynamics of the cool CGM of star forming galaxies in the Local Universe. 
\section{Discussion}\label{discussion}
We have seen in the previous Section that winds of cool clouds powered by the supernova explosions in the central galaxy are not a viable way to successfully describe the circumgalactic medium around star-forming galaxies. This result, in contrast with many claims of cool CGM gas being produced by outflows, both from observations \citep[e.g.][]{rubin14,schroetter19,martin19} and simulations \citep[e.g.][]{muratov15,ford16} may seem controversial, but is motivated by the unphysically high kinetic energy that these outflows would need to reproduce our dataset. In this Section we discuss the limitations of our models and we try to further verify our results relaxing some of the assumptions that we made in Section~\ref{model}. We will then describe the implications of our findings, especially regarding the origin of the cool CGM.
\subsection{Influence of the hot gas}\label{diffcor}
{ 
\renewcommand{\arraystretch}{1.5}
 \begin{table*}
\begin{center}
\begin{tabular}{*{6}{c}}
\hline  
\hline
 Model & $\log\ (m_{\rm{cl,start}}/\rm{M}_{\odot})$  & $\log\ (v_{\rm{kick}}/(\rm{km}\ \rm{s}^{-1}))$ & $\log \eta$ & $\theta$ & $f_{\rm{SN}}$\\
      &   &   &   & (degrees)   &\\
\hline 
Fiducial &$7.16\raisebox{.2ex}{$\substack{+0.71 \\ -0.63}$}$ & $2.73\raisebox{.2ex}{$\substack{+0.15 \\ -0.09}$}$ & $0.94\raisebox{.2ex}{$\substack{+0.41 \\ -0.38}$}$ & $79.43\raisebox{.2ex}{$\substack{+5.68 \\ -6.99}$}$ & $2.5$ \\
$\rm{M}_{\rm{cor}}=2\%\ \rm{M}_{\rm{bar}}$ &$7.00\raisebox{.2ex}{$\substack{+0.91 \\ -0.87}$}$ & $2.65\raisebox{.2ex}{$\substack{+0.06 \\ -0.06}$}$ & $0.42\raisebox{.2ex}{$\substack{+0.42 \\ -0.40}$}$ & $77.62\raisebox{.2ex}{$\substack{+7.49 \\ -11.56}$}$ & $0.6$ \\
Minor axis & $6.10\raisebox{.2ex}{$\substack{+1.93 \\ -0.59}$}$ & $3.02\raisebox{.2ex}{$\substack{+0.40 \\ -0.36}$}$ & $0.35\raisebox{.2ex}{$\substack{+0.92 \\ -0.23}$}$ & $53.70\raisebox{.2ex}{$\substack{+20.43 \\ -12.02}$}$ & $2.6$ \\
Inner regions & $5.78\raisebox{.2ex}{$\substack{+1.81 \\ -0.39}$}$ & $3.04\raisebox{.2ex}{$\substack{+0.46 \\ -0.48}$}$ & $0.42\raisebox{.2ex}{$\substack{+0.66 \\ -0.33}$}$ & $47.86\raisebox{.2ex}{$\substack{+19.75 \\ -12.38}$}$ & $3.3$\\
\hline
\end{tabular}
\end{center}
\captionsetup{justification=centering}
\caption[]{50th percentiles (with errors given by the 32nd and the 68th percentiles) of the posterior distributions of the four parameters obtained with the MCMC fits performed for the 4 models described in this work and consequent efficiencies of the supernova explosions.}\label{tab:testres}
   \end{table*}
   }
One of the main features of our models is the presence of a preexisting hot circumgalactic corona in the halos of galaxies, whose interaction with the cool clouds is strongly influencing their motion, as we have seen in Sections~\ref{model} and \ref{results}. The drag force \citep{marinacci11} acted by the hot gas decelerates the clouds, forcing them to have high initial velocities in order to reach the large distances where they are observed. Moreover, a fundamental effect of the hot gas is to pressure confine the cool clouds \citep{pezzulli19}. In fact, without the confinement of an ambient medium, the clouds would tend to expand and would not be in a stable state. We will indeed see later in this section that the density of the gas strongly influences the size of the cool CGM absorbers. As already mentioned in Section~\ref{intro}, the presence of hot gas at temperatures similar to the virial one and bearing a significant amount of baryons \citep[e.g.][]{shull12} is well justified by theoretical models \citep[e.g.][]{fukugita06} and has been confirmed by numerous observations \citep[e.g.][]{anderson11,li17,bogdan17,faerman20}.\\
Our description of this gas phase is physically motivated as a stratified medium in hydrostatic equilibrium with the dark matter halo and is consistent with the current evidence from observations, which are however still limited and mostly reliable only for the inner parts of the halos. We can see from Figure~\ref{fig:corona} how, in particular for the densities, there is a scatter of almost two orders of magnitudes between different observational estimates of the density of the hot gas. Our assumption of a mass of hot gas equal to 20\% of the total (cosmological baryon fraction) baryonic mass theoretically associated to the halo (see Section~\ref{hydro}) leads to densities that are well in agreement with the observational range. A slightly smaller mass fraction would however not be inconsistent with the observational estimates, given the large uncertainties. Since the efficiency of the drag force, and therefore the amount of deceleration of the clouds, depends on the density of the hot gas, the conclusion of the previous section might change using a corona with a mass lower than the one employed in our fiducial model.\\
We tested this possibility lowering the total mass of the hot phase to 2\% of the cosmological baryonic one: the density profiles for the three galaxy models of Table~\ref{tab:modprop} are showed in the top panel of Figure~\ref{fig:corona}. We can see how these profiles are already inconsistent with the majority of the data points, representing then a very extreme model for the hot gas. We exclude the possibility of having a corona with even lower masses.\\
In Table~\ref{tab:testres} we report the results of the MCMC analysis for a model with the properties explained above, in particular the median value of the posterior distributions of the 4 parameters, with the 32nd and 68th percentiles. We can see how the different mass of the corona changes the region of the parameter space recovered by the MCMC and therefore the best values of the parameters. As expected, with a more diffuse hot gas the clouds can more easily reach the observed distances from the central galaxies, where they are originated. The ejecting velocities and the mass of the clouds are however only slightly lower than in the previous case. At a fixed mass, the deceleration by drag force is indeed proportional to $\rho_{\rm{cor}}^{1/3}$ (see equation~\ref{eq:dragforce}). This dependence is therefore relatively weak and a variation in the normalization of the hot gas density does not dramatically influence the dynamics of the cloud. Moreover, this effect is counteracted by the larger cross section of the absorbers, since a lower gas density leads to larger clouds because of the pressure equilibrium. The main difference in the MCMC findings is in fact given by the mass loading factor, that is less than half than what previously found, since the larger clouds can more easily be intercepted by the lines of sight. We therefore need less clouds to reproduce the data, hence a lower mass loading factor.\\
The efficiency of energy transfer from the supernovae, calculated through equations~\eqref{eq:outpower} and~\eqref{eq:snpower} as in Section~\ref{model} and using the median values of the kick velocity and mass loading factor reported in Table~\ref{tab:testres}, is in this case equal to 60\%. This value means that only a fraction of the energy coming from the supernova explosions is being transferred as kinetic energy to the cool winds and is therefore more realistic than what we previously found with our fiducial model. However, as already mentioned, most of the SN energy is expected to be radiated away \citep{mckee77} and typical expectations for the value of the efficiency are of the order of 10\% or less \citep[e.g.][]{kim18,bacchini20}. The fraction that we find is therefore still far too high to be explained only with the stellar feedback as an energy source. Moreover, in this scenario the clouds have sizes even larger than in our fiducial case, with radii of more than 10 kpc. This feature makes the picture described in this section even more unrealistic than the one shown in Section~\ref{mcmc}. Most importantly, as already mentioned above, the mass of the hot gas used for this test is an extreme scenario, with densities barely consistent with a few and inconsistent with the majority of the observational estimates. We therefore conclude that the choice of the coronal medium does not influence the final result that cool supernova-driven outflows cannot reproduce the entirety of the cool circumgalactic gas.\\
A final important assumption that we made for the hot medium is about its kinematics. In particular, we assumed the corona to be static and supported only by the hydrostatic equilibrium with the dark matter halo. Theoretical studies \citep{pezzulli17} have shown however how the hot gas is required to rotate in order to satisfy galaxy evolution constraints. From an observational point of view the kinematic information are instead still limited, due to the spectral resolution, and mostly related only to the Milky Way \citep[e.g.][]{hodges16}. From both theory and observational estimates we expect the coronal gas to rotate at velocities lower than the rotational velocity of the galactic disc. Considering the very high outflow velocities that we have found for our best models, we do not expect the rotation of the corona to strongly influence the orbits of the cool clouds. We therefore believe that a rotating corona would not significantly influence our results, but we plan to explore this scenario in future works.\\
A general mechanism that is invoked to bring cold circumgalactic gas to very large distances is the entrainment of cold clouds, pushed by fast-moving hot winds to the external parts of the halos \citep[e.g.][]{gronke20}. In this scenario the corona is therefore not static as described by our models, but is flowing out from the central galaxy. The drag force from the outflowing hot gas would work in the opposite direction of the pre-existing hot gas employed in our models, accelerating the clouds and therefore lowering the values of the unphysical efficiencies discussed in Section~\ref{results}.
Including this additional push in our models could potentially decrease the initial ejection velocity of the clouds, towards more physically motivated scenarios. However, the gas in the hot phase should have masses comparable to the cool CGM and therefore the mass loading factor should substantially increase, increasing the necessary energy budget of the winds as a direct consequence. Moreover, the hot winds should show velocities of hundreds of $\rm{km}\ \rm{s}^{-1}$, which are not seen in the (uncertain) observational estimates of the corona kinematics \citep[e.g.][]{hodges16}, as well as metallicities close to the solar value, which are also in contrast with observational findings \citep[e.g.][]{bogdan17,bregman18}.
We therefore conclude that a scenario with hot gas outflowing and entraining cool clouds to the external regions is most likely not a viable description of the circumgalactic medium.
\subsection{Cool gas distribution}\label{45deg}
In this work we assumed that the cool circumgalactic gas is entirely part of outflows originated by the supernova feedback in the central galactic disc. However, different studies, both from a theoretical and an observational point of view \citep[e.g.][]{hafen19,martin19,schroetter19} have suggested that the cool clouds have two distinct origins and while some of them are part of galactic winds, others are instead longitudinally accreting from external gas towards the central galaxy. This picture results in a bimodal angular distribution of the absorbers, segregated along the minor (winds) and major axis (accretion) of the galaxy \citep[e.g.][]{kacprzak12}. As mentioned in Section~\ref{intro}, this scenario is still debated, in particular since recent studies find no correlation between the cool gas metallicity and the azimuthal angle of the absorptions \citep[e.g.][]{peroux16,pointon19}. Moreover, in our sample we do not see any evidence of a preferential axis in the detection of the cool CGM.\\ 
In the first part of this section we assume that the picture of the bimodal gas distribution is true and we calculate what would be the efficiency needed by galactic winds to reproduce only the observations along the galaxy minor axis.
In order to do this, we performed our MCMC analysis on a subsample of the observations, selecting only the lines of sight within 45 degrees from the galaxy minor axis (all the absorptions above the bisector in Figure~\ref{fig:Obslosplane}). In Table~\ref{tab:testres} we report the results of this analysis (minor axis). The models that best reproduce the observations have mass loading factors that are much smaller than the ones obtained with our fiducial model. This is expected since in this experiment the outflows need to reproduce a lower number of absorbers. However, the clouds are less massive than what previously predicted, forcing the ejection velocities to be extremely high, of the order of $1000\ \rm{km}\ \rm{s}^{-1}$. If we calculate the efficiency of the supernovae in transferring energy to the cool winds, we obtain $f_{\rm{SN}}\approx 2.6$, very similar to the one obtained through the comparison with all the absorbers. We therefore can conclude that supernova-driven outflows would need to be unphysically powerful in order to reproduce the observed cool gas kinematics along the galaxy minor axis. This could imply that there might be accretion of external gas (see Section~\ref{origin}) also along this axis, in agreement with what has been found by the studies of the gas metallicity \citep[e.g.][]{pointon19}.\\
A second viable way to reduce the energy budget required by the outflows is to assume that the cool gas does not reach the extremely large distances seen in the observations, comparable with the virial radii, but instead only the internal parts of the halos. The most external absorbers might have a different origin and not be related to the central galaxies (see Section~\ref{origin}).
To investigate further this hypothesis, we have performed a test similar to the one outlined above, but selecting only the lines of sight within a projected radius of 100 kpc (see Figure~\ref{fig:Obslosplane}). In this way, despite having less information to constrain our models (only 11 lines of sight satisfy the requirement), we can test whether the inner part of the cool CGM can be produced by galactic outflows in a physically motivated scenario. Note however that the distances of the observations are projected and therefore only a lower limit for the intrinsic distances of the clouds in our models. The results of this test might then be influenced by clouds that reach distances larger than 100 kpc.\\
In Table~\ref{tab:testres} we show the best fit results of this analysis (inner regions). We can see how the main differences with the previous parameter values are the mass of the clouds and the aperture of the outflowing cone. A mass of less than $10^6\ \rm{M}_{\odot}$ is in this case sufficient for the clouds to reach the distances of this subsample of the data and the outflows do not need to be isotropic ($\theta\lesssim50^{\circ}$), contrary to what we found for our fiducial model. Nevertheless, the ejection velocities and mass loading factors lead to supernova efficiencies that are still extremely high, even higher than the previous ones. This result is then opposite to what we expected (lower efficiencies), but it is motivated by the fact that these models reproduce better than the fiducial model the gas kinematics observed in the inner regions (a Kolmogorov-Smirnov test between the model and observed velocity distributions for these 11 lines of sight gives a probability value $p=0.13$ for the fiducial model and $p=0.26$ for the model described in this section). The findings of this experiment demonstrate that most of the observations up to $\sim100$ kpc from the central galaxies cannot be explained only with galactic winds, as generally predicted by theoretical works \citep[e.g.][]{brook12,ford14}. Indeed, often simulations employ efficiencies of energy transfer that are larger than unity to produce such large-scale outflows, which is consistent with our findings and demonstrates that these galactic outflows are not viable to describe the circumgalactic medium, except for maybe the gas residing in the very central regions of the halos. This result is also in line with the study of \cite{concas17}, who found that high-velocity galactic winds are rare in the Local Universe and, when present, seem to be driven mainly by AGN activity, rather than pure star formation.
We conclude then that most of the cool CGM is not formed by galactic wind outflows, but has instead a different origin, as we will discuss more in detail in Section~\ref{origin}.
\subsection{Limitations of the model}\label{dischydro}
The approach of this work is to use semi-analytical models to study the CGM of star-forming galaxies. As we explained in Section~\ref{intro}, our method has several advantages with respect to hydrodynamical simulations, both cosmological and idealized high-resolution ones. The former are indeed able to track the whole extent of the circumgalactic gas inside a galaxy halo but are not able to resolve the cool clouds and their interactions with the hot medium; the latter instead can trace the hydrodynamic instabilities and thermal conduction that characterize the cloud evolution, but can only focus on a very small part of the halo. Our parametric approach allows us to have a comprehensive view of the CGM and, through a thorough comparison with the observations, to derive some fundamental properties of this medium. This comes however with some limitations, that we will discuss in the following.\\
Our representation of the cool clouds is simplified, since the real clumps of cool gas are not spherical, but have irregular shapes due to the second-order effects taking place at the interface between hot and cool gas, as clearly shown by hydrodynamical simulations. Physical effects like the Kelvin-Helmoltz (KH) instability, that tend to strip gas from the clouds hence destroying them, are neglected in our parametrization, as well as thermal conduction, which on the one hand suppresses the KH instability and on the other makes the clouds evaporate into the hot medium \citep[e.g.][]{ armillotta17,lan18}. 
All these effects influence the survival time of the cool clouds and can lead to their evaporation into the corona, which is a major feature that is not taken into account in our models. \cite{lan18}, using semi-analytical models similar to the one developed here, found indeed that more than $50\%$ of the cool clouds ejected from the central galaxy will evaporate into the hot halo. Nevertheless, the disruption of the clouds goes in the direction of strengthening our result, since the clouds will lose mass and therefore would need even higher velocities than the one predicted by our model in order to both overcome the drag force deceleration of the hot gas and reach the external regions of the halo before disruption. We then conclude that the cloud evaporation would not change the findings of this work.\\
Different studies \citep[e.g.][]{marinacci11,armillotta16,gronke18} have found how, in the region close to the galaxy disc, the gas at the interface between cold clouds and the hot corona reaches temperatures that allows it to cool very rapidly and therefore to condensate, forming more cold gas. This has been proposed as a viable mechanism to accrete new gas and sustain the star formation of the central galaxy \citep{fraternali08}, or to explain the survival of cold gas clumps embedded in fast hot gas outflows \citep{schneider18}. This is however restricted to the very internal regions of the halos, a few kpc above the galactic disc, where the density and temperature of the corona create the condition for the condensation of the mixed gas, while the evaporation of the clouds, whose implications for our findings have already been discussed, seems to be the most relevant effect at larger distances \citep[e.g.][]{armillotta17}, which are the ones we are probing with our study.\\
Another limitation of our models is to consider that the outflows are powered only by the supernova feedback, neglecting in particular a possible central AGN, which is often invoked in hydrodynamical simulations \citep[e.g.][]{nelson19} as a source of the circumglactic gas. Although we have no clear evidence in our galaxies of AGN activity \citep{werk13,borthakur15}, previous AGN outflows could have influenced the motion of the gas and could have been a potential additional source of energy for the cool winds. This would therefore go into the direction of making the outflow scenarios more realistic from an energetic point of view. We however consider it unlikely for the AGN to be a dominant energy source for the cool outflows along the whole galaxy lifetime and we conclude that this addition would most likely not influence our main result.
\subsection{Origin of the cool CGM}\label{origin}
Throughout this paper, we have proved that winds from the galactic disc, powered by the supernova feedback, cannot account for most of the cool circumgalactic gas around star-forming galaxies. In this Section we will speculate on the possible origin of this gas.
\subsubsection{External origin}\label{external}
The main reason for the unphysical efficiencies that we found in this work is the fact that the cool clouds have to travel from the galactic disc out to very large distances in order to reproduce the observations. Zoom-in hydrodynamical simulations \citep[e.g.][]{oppenheimer18,nelson19} show however how the circumgalactic gas is probably produced by a combination of different processes and while the galactic winds have a crucial role in the gas dynamics and origin, a significant part of it is probably accreted from the intergalactic medium. Cosmological models predict indeed that galaxy halos are acquiring external gas \citep[e.g.][]{fakhouri10} and this accretion might be the origin of the cool absorbers observed by the COS surveys. In particular, we could expect filaments of low-metallicity gas \citep{lehner16} to enter the galactic virial radius and to be disrupted into clouds by the interactions with the hot pre-existent CGM. A similar scenario has also been proposed for the cool gas around early-type galaxies, where outflows from the passive central galaxy likely play a minor role in the origin of the circumgalactic gas \citep[e.g.][]{huang16}. In the halos of these galaxies, the drag force acted by the corona on the infalling clouds slows down their motion and leads to cloud velocities that are in agreement with the observations \citep{zahedy19}.\\ 
Since we have proven that galactic outflows fail in successfully describing the cool CGM of star-forming galaxies, the accretion of intergalactic gas might be a more plausible scenario for the origin of the cool gas. This picture of inflow of external gas would be consistent with the cold-mode accretion scenario \citep[e.g.][]{keres09}, where cold gas filaments directly enter the galactic virial radius and can eventually reach and feed the central galaxy with new gas \citep{dekel09}. As we have seen in Section~\ref{45deg}, our work leads to the conclusion that the outflows can be responsible for the cool CGM only in the very central regions of the halo. We conclude that the most likely origin of the majority of the cool circumgalactic clouds is instead the accretion of external gas, as we have found for early-type galaxies  \citep{afruni19}. As already mentioned, this scenario would also be consistent with the fact that a significant part of the cool gas absorbers observed around low-redshift galaxies have low metallicities \citep[e.g.][]{wotta19}.\\
A second possible external origin for the cool CGM is the stripping of gas from satellite galaxies. Recent simulations \citep[e.g.][]{alcazar17,hafen19} have found that winds of cool gas from satellites could produce a significant fraction of the cirgumgalactic medium of the central object. Moreover, lately there has been evidence of similarities between the kinematics and positions of dwarf galaxies around M31 and the properties of the cool ionized gas \citep[see][]{lehner20p} observed in its halo. Even though we have neglected the stripping from satellite galaxies in our work, this origin can not be ruled out and is likely to be part of the complex picture of the CGM of star-forming galaxies.
\subsubsection{Condensation of the hot gas}
We have seen that the accretion of external gas is one of the most likely scenarios for the origin of the cool CGM. Another way to form cold gas is however through the cooling of the corona residing in the galaxy halo. The direct radiative cooling of the hot CGM (hot-mode accretion) is not an efficient way to create cold gas, since its very low densities, especially in the outermost regions, imply cooling times of the order of the Hubble time, but thermal instabilities could accelerate  this process, forming cold clumps that eventually fall down towards the galaxies. This type of condensation (still debated, see \citealt{binney09}) has been recently predicted by some theoretical models, where the feedback from the central galaxy regulates the development of the instabilities \citep[e.g.][]{thompson16,voit18}, which however generally originate the cool clouds at smaller radii than the typical scales of the CGM analysed in this paper. Recent hydrodynamical simulations \citep{nelson20} also show that, at least for massive elliptical galaxies, a not negligible amount of cool gas can originate from thermal instabilities triggered by density perturbations in the hot corona.\\
Even though we believe that most of the cool gas has an external origin (see Section~\ref{external}), the condensation of the hot medium residing in the galaxy halos cannot be excluded and could play a role in the creation of part of the cool CGM.
\section{Summary and conclusions}\label{conclusions}
In this work we have developed parametric semi-analytic models in order to describe the kinematics of the cool circumgalactic medium observed in the halos of low-redshift star-forming galaxies.\\
In particular, we have made the main assumption that the cool gas absorbers are part of galactic wind outflows powered by the supernova explosions in the galaxy disc. We have investigated whether this kind of scenario is able to reproduce the observational dataset of the COS-Halos and COS-GASS surveys, focusing on a subsample of 41 star-forming galaxies. We have compared the outputs of our models with the observations using a bayesian MCMC analysis, in order to constrain the main properties of the cool gas outflows.\\ 
Through this analysis we have obtained the following results:
\begin{enumerate}
\item axisymmetric models of cool clouds outflowing from the galactic disc are able to reproduce the line-of-sight velocities and number of components observed by the COS-Halos and COS-GASS surveys. In order for the outputs of our models to be consistent with the observations, the clouds need to have masses of about $10^7\ \rm{M}_{\odot}$ and initial ejection velocities higher than $500\ \rm{km}\ \rm{s}^{-1}$, while the outflows need to be almost isotropic and with mass loading factors of about 10;
\item given the ejection velocities and mass loading factors necessary to reproduce the data, the energy required by these cool outflows is much larger than the one available from supernova explosions. We can therefore conclude that cool outflows from stellar feedback are not the origin of the cool CGM;
\item since we excluded the outflows as a source for the cool gas, we conclude that most of the this medium has different origins, the most promising being accretion from the intergalactic medium, but also potentially condensation of the hot halo gas and gas stripping from satellite galaxies.
\end{enumerate}
With this paper we reinforced the idea that most of the cool circumgalactic medium is originated by the accretion of gas from the IGM, similar to what we have previously found for massive early-type galaxies in \cite{afruni19}. Our main conclusion is therefore that supernova-driven outflows from the central galaxies most likely have a much smaller role in the dynamics and origin of the cool CGM than what is generally believed. 

\section*{Acknowledgements}
We thank an anonymous referee for the helpful comments and useful suggestions. We are also grateful to Pavel Mancera Piña, Cecilia Bacchini and Asger Gr{\o}nnow for useful discussions that helped improve the quality of this work. AA would like to thank Jo Bovy for the help in the modification of the GALPY package. GP acknowledges support by the Swiss National Science Foundation, grant PP00P2\_163824 and from the Netherlands Research School for Astronomy (NOVA).

\section*{Data Availability}
The data underlying this article are available in the article and in its online supplementary material.

\vspace{-0.3cm}




\bibliographystyle{mnras}
\bibliography{biblio} 



\vspace{-0.4cm}

\appendix

\section{Fitting of the galaxy images}\label{galfit}
\begin{figure*}
   \includegraphics[width=1\linewidth]{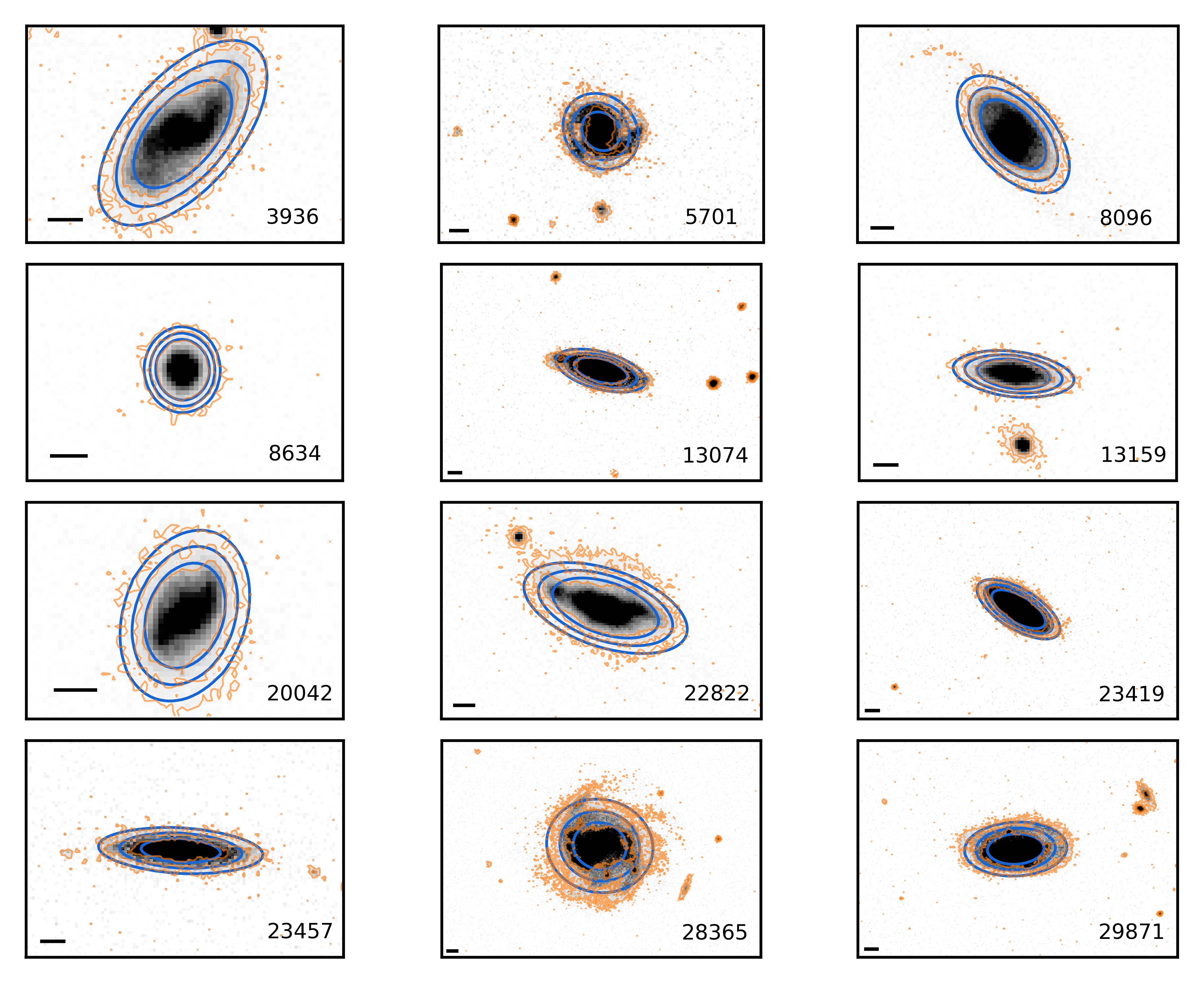}
   \caption{SDSS images of 12 out of the 24 galaxies selected from the COS-GASS survey. The contours are placed at 2, 4 and 8 times the noise level in each image. The orange contours represent the data, while the blue ones represent the GALFIT best model. The black bar depicts an angular size of 3 arcseconds.}
              \label{fig:galfit1}%
    \end{figure*}

\begin{figure*}
   \includegraphics[width=1\linewidth]{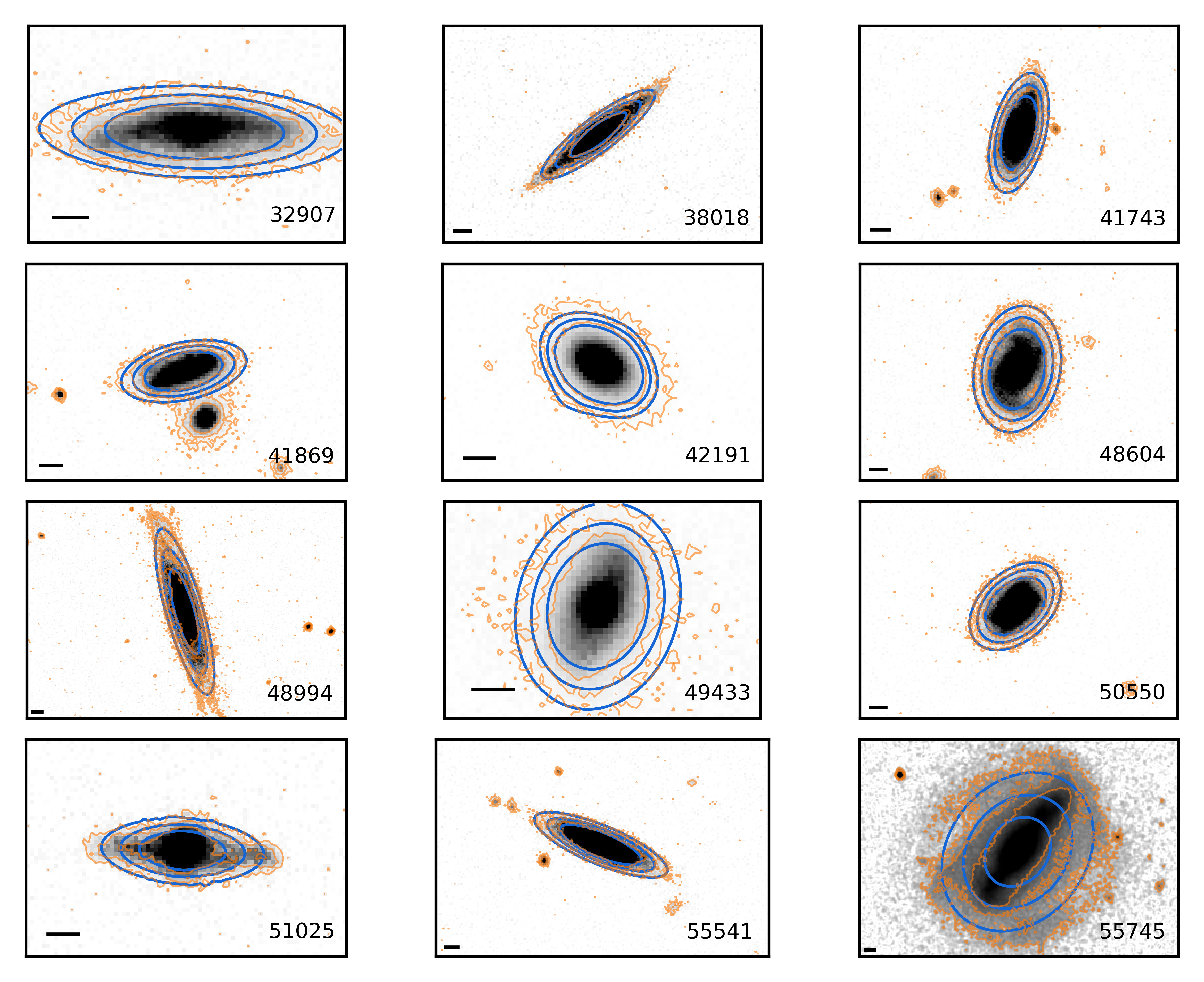}
   \caption{Same as Figure~\ref{fig:galfit1}, for the rest of the galaxies in our sample of COS-GASS galaxies.}
              \label{fig:galfit2}%
    \end{figure*}

\begin{figure*}
   \includegraphics[width=1\linewidth]{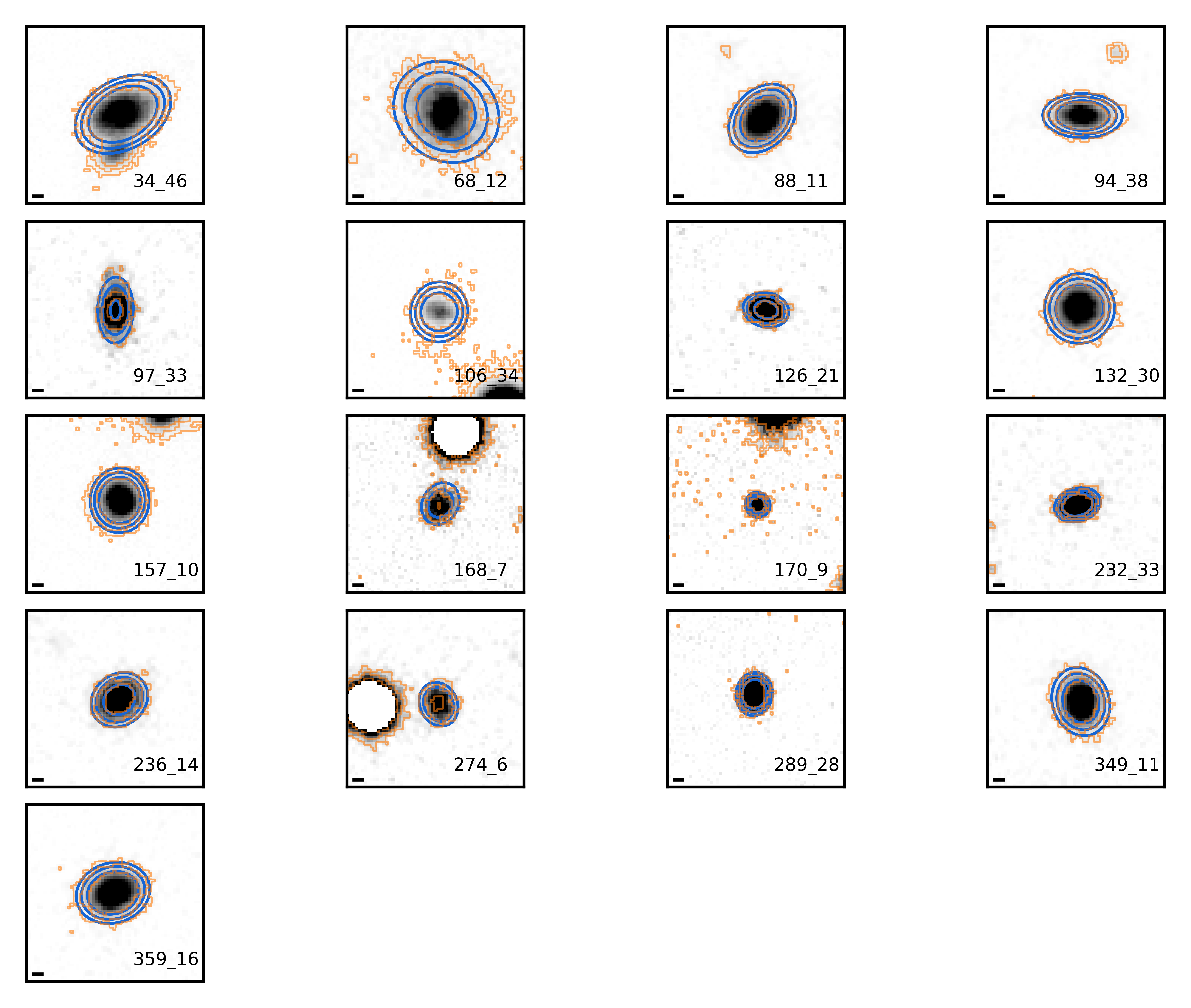}
   \caption{DESI legacy images of the 17 galaxies selected from the COS-Halos survey. The countours and the black bar are the same as in Figures~\ref{fig:galfit1} and \ref{fig:galfit2}.}
              \label{fig:galfit3}%
    \end{figure*}

Some of the parameters of the galaxies in our samples, reported in Table~\ref{tab:galprop}, have been retrieved performing a fitting of the data with the software GALFIT, which uses parametric functions to model objects in 2-dimensional images \citep[see for more details][]{peng10}. The images of the 41 galaxies in our sample have been retrieved from the Sloan Digital Sky Survey (SDSS, \citealt{blanton17}) for the COS-GASS galaxies (24/41 of our sample) and from the DESI legacy imaging survey \citep{dey19} for the COS-Halos galaxies (17/41 of our sample) and are shown in Figures~\ref{fig:galfit1}, \ref{fig:galfit2} and \ref{fig:galfit3}.\\
In particular, we used the GALFIT software to fit an exponential disc parametric function on the galaxies, described by
\begin{ceqn}
\begin{equation}
\Sigma(r)=\Sigma_0 \exp{-R/R_{\rm{\rm{d}}}} ,
\end{equation}
\end{ceqn}
where $\Sigma_0$ is the central surface brightness and $R_{\rm{d}}$ is the disc length. 
The best models resulting from the fits are shown in Figures~\ref{fig:galfit1}, \ref{fig:galfit2} and \ref{fig:galfit3}, where we can see how the model contours (blue) match very well the observational ones (in orange). The contour levels are placed at 2, 4 and 8 times the noise level in each of the images.\\
In particular, for our models (see Section~\ref{model}) we are interested in the galaxy inclinations, the positions of the sightlines with respect to the galaxy disc and the galaxy disc scale length. The latter is directly inferred as one of the parameters of the GALFIT fitting\footnote{All the distances obtained from GALFIT are transformed from pixel to arcsec using the pixel scales of the SDSS and legacy surveys, respectively 0.396 and 0.06. They are subsequently transformed in kpc using the \textit{cosmology.FlatLambdaCDM} routine of the python \textit{astropy} package.}, while the other two are calculated starting from the galaxy axis ratio and position angle, which are the direct outputs of GALFIT.\\
The inclinations in particular are obtained through the formula 
\begin{ceqn}
\begin{equation}
\cos{i}=\sqrt{\frac{(b/a)^2-q_0^2}{1-q_0^2}}\ ,
\end{equation}
\end{ceqn}
where $b$ and $a$ are respectively the minor and major axis of the disc, while $b/a$ and $q_0=0.2$ are respectively the apparent and intrinsic disc axis ratio. The position angle is instead used together with the information of the positions of the quasar sightlines and the galaxies \citep[from][]{tumlinson13,borthakur15} to calculate the projected distances of the quasars from the galaxies, in the reference frame of the galactic disc.\\
The inclinations, distances of the lines of sight and disc scale lengths of the 41 galaxies in our sample are reported in Table~\ref{tab:galprop}.
\section{Tests on mock data}\label{artdata}
\begin{figure*}
   \centering
   \includegraphics[clip, trim={0 0 0cm 0}, width=10.cm]{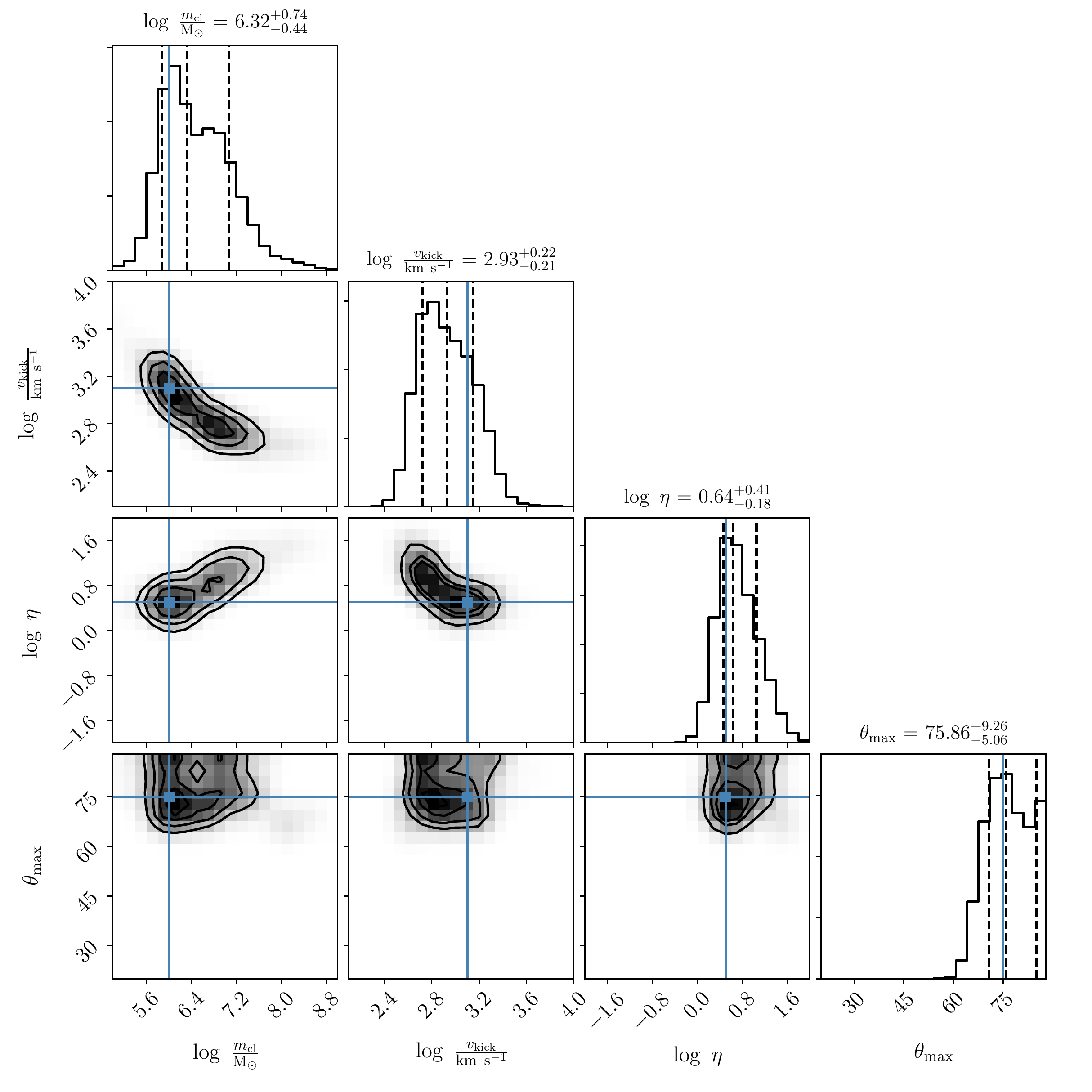}
   \caption{Corner plot with the MCMC results, representing the one and two dimensional projections of the posterior probabilities for the four free parameters of our model. Test on the mock data obtained with input values of the parameters $\log m_{\rm{cl}}=6$, $\log v_{\rm{kick}}=3.1$, $\log \eta=0.5$ and $\theta_{\rm{max}}=75^{\circ}$, represented by a blue dot in the parameter space.}
              \label{fig:test1}%
    \end{figure*}

\begin{figure*}
   \centering
   \includegraphics[clip, trim={0 0 0cm 0}, width=10.cm]{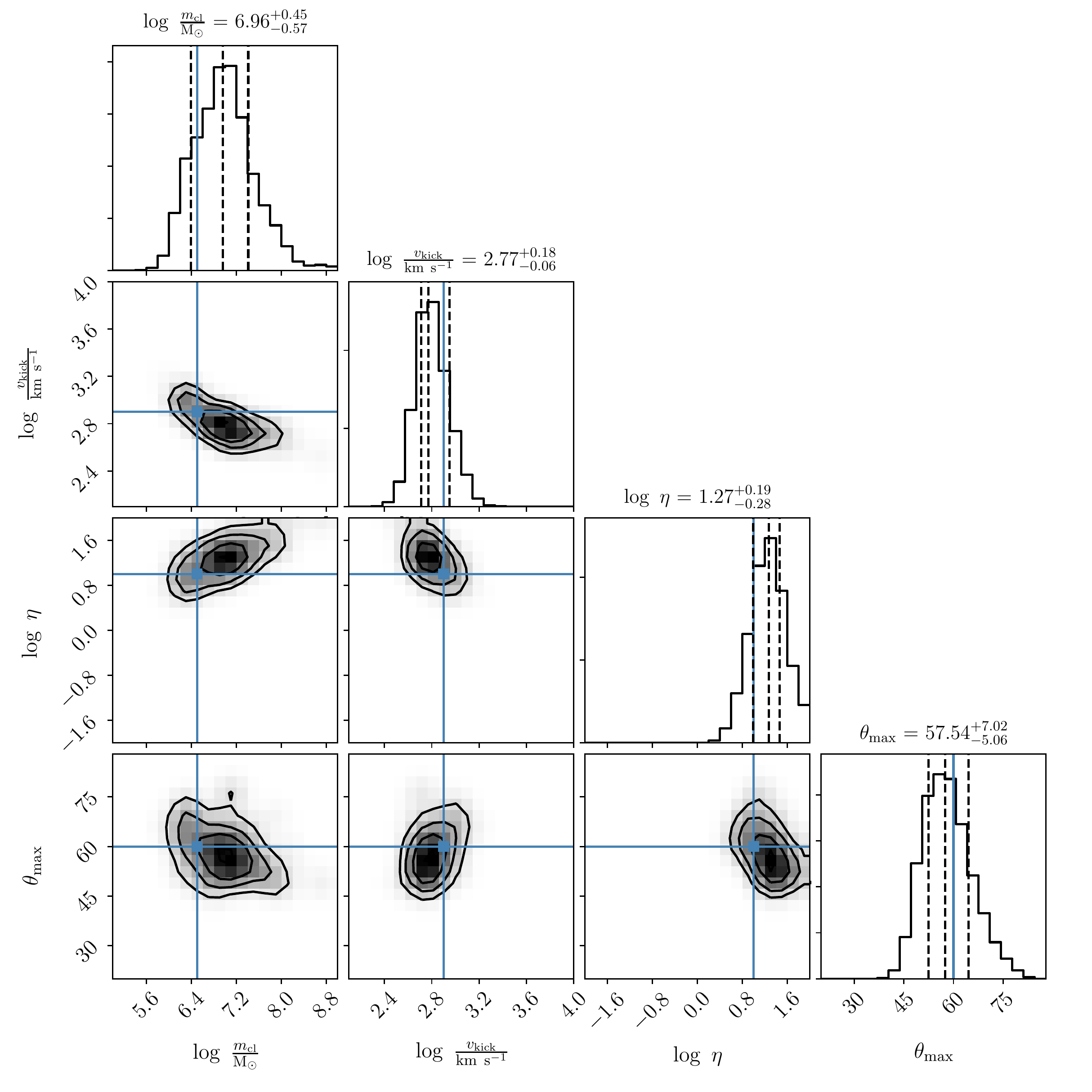}
   \caption{Same as Figure~\ref{fig:test1}, for the mock data obtained with $\log m_{\rm{cl}}=6.5$, $\log v_{\rm{kick}}=2.9$, $\log \eta=1$ and $\theta_{\rm{max}}=60^{\circ}$.}
              \label{fig:test2}%
    \end{figure*}

\begin{figure*}
   \centering
   \includegraphics[clip, trim={0 0 0cm 0}, width=10.cm]{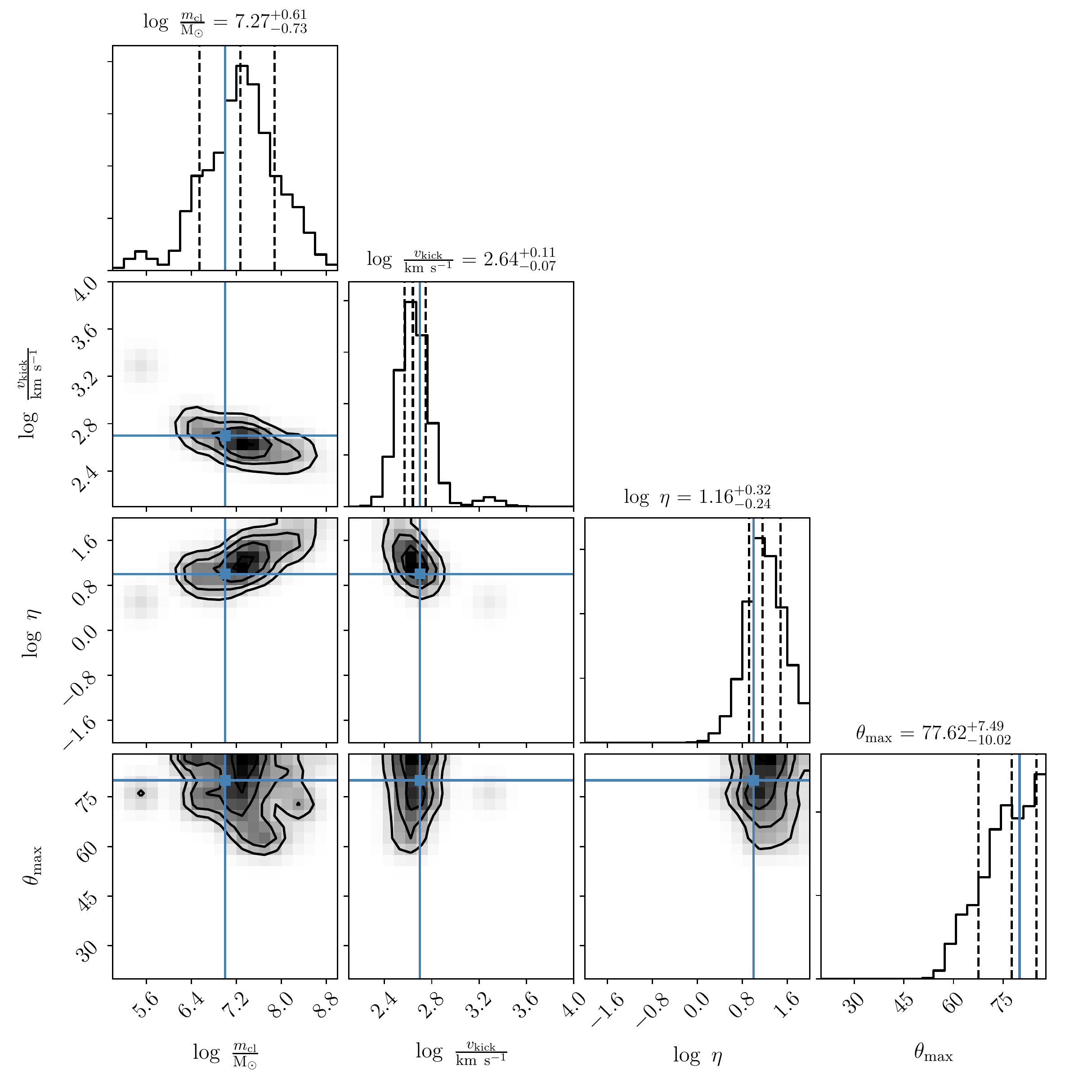}
   \caption{Same as Figures~\ref{fig:test1} and \ref{fig:test2}, for the mock data obtained with $\log m_{\rm{cl}}=7$, $\log v_{\rm{kick}}=2.7$, $\log \eta=1$ and $\theta_{\rm{max}}=80^{\circ}$.}
              \label{fig:test3}%
    \end{figure*}

In this appendix we present tests that we carried out to investigate the validity of the likelihood described in Section~\ref{likelihood}. In particular, we applied the MCMC analysis on different sets of mock data, with the purpose of verifying whether with our analysis we can retrieve the true values of the 4 free parameters of our models: $m_{\rm{cl}}$, $v_{\rm{kick}}$, $\eta$ and $\theta_{\rm{max}}$, under the assumption that the cool CGM is outflowing from the central galaxies, powered by the supernova explosions in the disc.\\
We considered specific choices of these four parameters and through the synthetic observations described in Section~\ref{synthetic} we created the mock observations that correspond to a certain number of velocity components for each line of sight. We then performed the MCMC analysis using the likelihood described in Section~\ref{likelihood} and treating the mocks as a set of observations. We present here 3 different tests that we carried out using 3 different choices of the input parameters. In Figures~\ref{fig:test1}, \ref{fig:test2} and \ref{fig:test3} we show the results of these tests, showing in particular the one and two dimensional posterior distributions of the free parameters, with the blue points representing the values of the parameters from which the mock data have been created. We can see how these values, in all the tests, have been successfully retrieved by our MCMC analysis. From the corner plots we observe some degeneracy, meaning that there is not a single choice of parameters, but an entire region of the parameter space that leads to models that have outputs consistent with the real one. The real value is however always within the 32nd and 68th percentiles of the posterior distributions.\\
From this analysis we conclude that, if the cool clouds are part of galactic outflows, with our likelihood we are able to properly find the choice of parameters, or at least the region of the parameter space, that best reproduces the real observational data.

\bsp	
\label{lastpage}
\end{document}